\documentclass[conf]{new-aiaa}
\usepackage{multicol}
\usepackage{graphicx}
\usepackage{afterpage}

\usepackage{epsfig,wrapfig}
\usepackage{amstext,amsfonts,amsmath,amscd, amssymb}
\usepackage[font=small,skip=0pt]{subcaption}
\usepackage{psfrag}
\usepackage{epstopdf}
\usepackage{color}
\usepackage{mdwlist}
\usepackage{wrapfig}
\usepackage{todonotes}
\usepackage{comment}
\usepackage[percent]{overpic}
\usepackage{setspace}
\usepackage{array}

\definecolor{green1}{RGB}{221,242,151}
\definecolor{green2}{RGB}{30,150,30}
\definecolor{purple1}{RGB}{149,87,164}
\definecolor{blue1}{RGB}{42,144,158}
\definecolor{dblue}{RGB}{11,48,84}
\definecolor{lblue}{RGB}{150,150,250}
\definecolor{mygray}{gray}{0.6}

\graphicspath{{./figures/}}

\newcommand{\overbar}[1]{\mkern 1.5mu\overline{\mkern-1.5mu#1\mkern-1.5mu}\mkern 1.5mu}

\def\pd{\partial}

\newcommand{\mcal}[1]{\mathcal{#1}}

\newcommand{\commentout}[1]{}

\def\la{\langle}
\def\ra{\rangle}

\title{\textbf{Towards a Predictive Hybrid RANS/LES Framework}}
\author{Sigfried Haering\footnote{Postdoctoral Researcher, Institute for Computational Engineering and Sciences}, 
Todd A.~Oliver\footnote{Research Scientist, Institute for Computational Engineering and Sciences},
Robert D.~Moser\footnote{Professor, Institute for Computational Engineering and Sciences and Department of Mechanical Engineering}}
\affil{The University of Texas at Austin, Austin, Texas, 78712}

\date{}

\begin{document}

\maketitle

\begin{abstract}
Predictive simulation of many
complex flows requires moving beyond Reynolds-averaged
Navier-Stokes (RANS) based models to representations resolving at
least some scales of turbulence in at least some regions of the flow.
To resolve turbulence where necessary while avoiding the cost of performing large eddy
simulation (LES) everywhere, a broad range of hybrid RANS/LES methods
have been developed.  While successful in some situations, existing
methods exhibit a number of deficiencies which limit their predictive
capability in many cases of interest, for instance in many flows
involving smooth wall separation.  These deficiencies result from
inappropriate blending approaches and resulting inconsistencies
between the resolved and modeled turbulence as well as errors
inherited from the underlying RANS and LES models.  This work details
these problems and their effects in hybrid simulations, and develops a
modeling paradigm aimed at overcoming these challenges.  The new
approach discards typical blending approaches in favor of a
hybridization strategy in which the RANS and LES model components act
through separate models formulated using the mean and fluctuating
velocity, respectively.  Further, a forcing approach in which
fluctuating content is actively transferred from the modeled to the
resolved scales is introduced.  Finally, the model makes use of an
anisotropic LES model that is intended to represent the effects of
grid anisotropy.  The model is demonstrated on fully-developed,
incompressible channel flow and shown to be very promising.
\end{abstract}

\section{Introduction}
\label{sec:intro}

It is widely recognized that combining Reynolds-averaged Navier-Stokes
(RANS) models with large eddy simulation (LES) in a hybrid modeling
framework can, in principle, provide a computationally tractable,
predictive tool for simulation of high Reynolds number flows in
complex geometries.  Since the 1990s, many hybrid simulation
approaches, including Detached Eddy Simulation (DES) and
variants~\cite{spal:2006,shur:2008,riou:2008,grits:2012,jee:2012},
partially-averaged Navier-Stokes~\cite{giri:2006, Foroutan2014},
partially-integrated transport models~\cite{Chaouat2005, Chaouat2017},
scale-adaptive simulation~\cite{Menter2005, Menter2010}, zonal
RANS/LES approaches~\cite{quem:2005,deck:2005}, and
others~\cite{froh:2008,bhus:2012,jenn:2017}, have been developed.  However, despite this
research and the intuitive appeal of the ideas underlying hybrid
models, because of a host of issues introduced by the blending of RANS
and LES models as well as deficiencies inherited from the underlying
models, a hybrid approach capable of reliably predicting a wide range
of flows does not yet exist.  The overall objective of this work is to
develop such a predictive hybrid RANS/LES modeling framework.  This
paper makes two contributions towards this goal.  First, it highlights
a set of deficiencies common to many hybrid methods.  Second, it
introduces a general modeling framework in which these deficiencies
can be addressed in a systematic fashion.

The deficiencies of existing methods come in three categories.  The first
is related to combining RANS and LES.  Naturally, the hybridization of
RANS and LES is at the heart of hybrid models.  However, in most
existing methods, the form of the transition between RANS and LES-like 
modeled stress is ad-hoc.  The deficiencies of such blending lead to non-physical behavior of the 
blended state, thereby corrupting the hybrid simulation.  As discussed
in~\S\ref{sec:blending_problems}, it appears a primary cause of these
shortcomings is the blending of RANS and LES through a single stress
model, which is invalid because the requirements and characteristics
for the two models are incompatible.  Instead, an alternative
hybridization strategy is required.


Second, in complex hybrid simulations, fluid inevitably flows from
RANS to LES regions and vice versa and between regions of varying LES
resolution. These situations require that energy associated with
turbulent fluctuations be exchanged between resolved and modeled
scales.  Most current methods affect this exchange passively,
relying on natural instabilities and dissipation.  However, as shown
in~\S\ref{sec:forcing_motivation}, to make effective use of LES-like
resolution while avoiding the well-known modeled stress depletion (MSD) problem~\cite{spal:2009}, it is
necessary that the exchange between modeled and resolved turbulence be
actively managed or forced.

Finally, in LES regions, hybrid models inherit the deficiencies of the
subgrid stress (SGS) model formulations they are based upon.  Typical
LES models are built assuming isotropic unresolved turbulence and
homogeneous, isotropic filtering/resolution.  Most SGS models do not
consider that these assumptions may be violated, leading to poor
performance~\cite{spal:2009}.  In hybrid applications, these
shortcomings can be particularly acute because the grids usually have
coarse LES resolution as well as anisotropic cells and inhomogeneous
resolution.  On such grids, the unresolved scales contribute a
significant portion of the total turbulent stress and are still
strongly anisotropic.

Overcoming these shortcomings necessitates a three-pronged approach.
First, to address issues associated with blending, this work
introduces a hybridization technique in which RANS and LES SGS models
act through separate model formulations using the mean and fluctuating
velocity fields, respectively.  Second, the approach is equipped with
a forcing mechanism that affects the exchange between resolved and
modeled turbulence on time scales that allow the resolved turbulence
to maintain local equilibrium, thereby avoiding the need to generate
realistic synthetic turbulence.  Third, the approach is able
to make use of nearly any underlying RANS and LES models, allowing the
inadequacies of typical models to be overcome as improved models are
developed.  For example, in this work, an anisotropic LES model that
represents the effects of grid anisotropy is used.  Taken together,
these three developments are intended to resolve the problems noted
previously.

The approach is demonstrated on fully-developed, incompressible,
turbulent channel flow at $Re_{\tau} = 5200$.  The results show that
the mean velocity is well predicted for a range of levels of resolved
turbulence and that the turbulent kinetic energy is better predicted
than in RANS.  Thus, the new method appears promising, although
further testing on a broader range of cases is of course necessary.

The remainder of the paper begins with a more in depth examination of
the problems briefly discussed above in~\S\ref{sec:background}.
Then, \S\ref{sec:modeling} describes the approach developed here,
and \S\ref{sec:results} shows preliminary results for channel flow.
Finally,~\S\ref{sec:conclusions} provides conclusions.

\section{Motivation: Requirements for Predictive Hybrid RANS/LES}
\label{sec:background}
As noted in~\S\ref{sec:intro}, there are three major issues in
existing hybrid models.  This section describes these issues in more
detail, including examples of the errors they lead to in practical
simulations.

\subsection{Inadequacy of Blending} \label{sec:blending_problems}
The RANS and filtered Navier-Stokes equations are formally similar, and
common RANS and LES SGS models also take a similar form, where the
effects of unresolved turbulence are represented through an eddy
viscosity.  These similarities lead to the alluring idea of blending the
model stress such that the simulation transitions from pure RANS to full
LES as a function of some measure of the available resolution.
However, closure models built for RANS and LES have conflicting
characteristics that make it difficult, if not impossible, to
successfully blend them through a single eddy-viscosity-based stress
model.  There are two primary issues.  First, typical RANS transport
equations are not designed for use with fluctuating state.  Second,
typical LES models are not designed to simultaneously predict the
subgrid contribution to the Reynolds stress and the dissipation.

\subsubsection{RANS model behavior with fluctuating state}
One problem is that transport equation models built for use in RANS
may severely misbehave when applied to LES fluctuations.  For example,
consider the destruction term in the standard model for the
evolution of the turbulent dissipation rate $\varepsilon$:
$\mathcal{D}_{\varepsilon} = C_{\varepsilon2} \varepsilon^2 / k$.  As an
approximation of the destruction term in the exact $\varepsilon$
equation, this form has little physical justification.  Nonetheless, for RANS,
the model can be collectively calibrated to behave reasonably well.  However, 
in an LES, the fluctuations have a substantial impact on the term's 
mean value.  To illustrate the impact of the fluctuations, one can expand 
$1/k$ in a Taylor series about $1/\langle k \rangle$.  Retaining the linear and quadratic terms in $k'$ leads to the following approximation:
\begin{align*}
\left\langle{} \frac{\varepsilon^2}{k} \right\rangle{}
& \approx
\frac{ \langle{} \varepsilon \rangle{}^2}{\langle k \rangle}
\left( 
1
+
\frac{ \langle{} \varepsilon'^2 \rangle{} }{\langle \varepsilon \rangle^2}
+
\frac{ \langle k'^2 \rangle }{\langle k \rangle^2}
-
\frac{ \langle \varepsilon' k' \rangle }{\langle \varepsilon \rangle \langle k \rangle}
-
\frac{ \langle{} \varepsilon'^2 k' \rangle }{\langle \varepsilon \rangle^2 \langle k \rangle}
+
\frac{ \langle{} \varepsilon' k'^2 \rangle{} }{\langle \varepsilon \rangle \langle k \rangle^2}
+
\frac{ \langle{} \varepsilon'^2 k'^2 \rangle{} }{\langle \varepsilon \rangle^2 \langle k \rangle^2}
\right).
\end{align*}
While it is not obvious what correlations may arise from all the terms, the variances in $k'$ 
and $\varepsilon'$ are clearly positive definite with fluctuations present.  
Due to non-zero variances and covariances when $k$ and $\varepsilon$ 
are fluctuating,  $\la{}\mathcal{D}_\varepsilon\ra$ is not simply 
$C_{\varepsilon2} \langle{}\varepsilon\rangle^2/\langle{}k\rangle$ as it is in RANS.
Instead, the behavior of the entire system of equations is altered 
depending on how much turbulence is resolved.

\begin{figure}[htp]
\begin{center}
\includegraphics[width=0.5\linewidth]{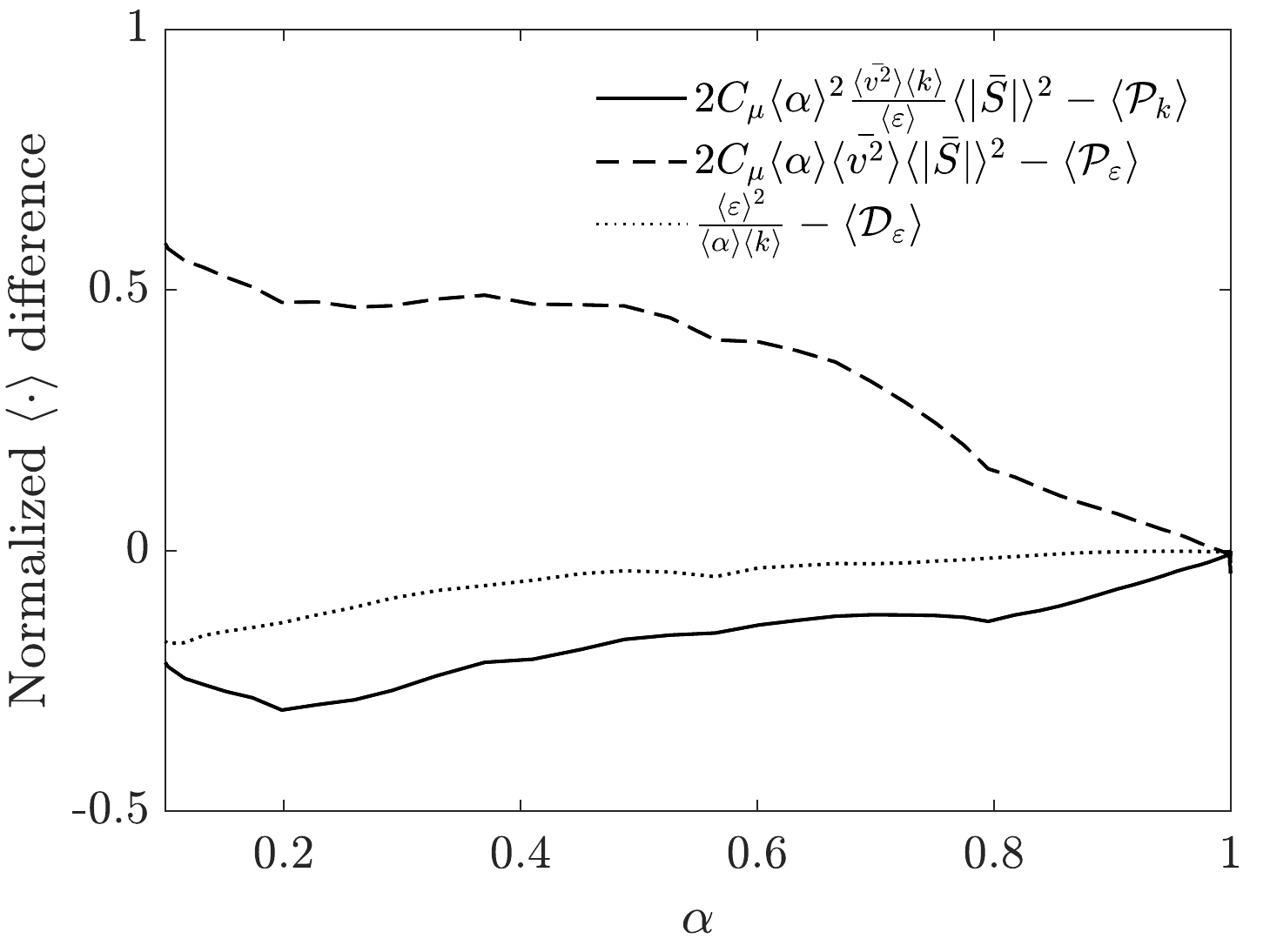}
\end{center}
\caption{ Differences in source terms for $k$ and $\varepsilon$ as
  calculated from the means of the input quantities and the mean of
  the full fluctuating term.
  The production of
  $k$ is normalized by $\langle\mathcal{P}_k\rangle$ while both
  production and destruction of $\varepsilon$ are normalized
  $\langle\mathcal{D}_\varepsilon\rangle$.}
\label{fig:chan_expected_sources}
\end{figure}

\begin{figure}[htp]
\begin{center}
\begin{subfigure}[b]{0.49\linewidth}
   \includegraphics[trim={1.25cm 0cm 1.5cm 0cm},clip=true,width=0.99\linewidth]{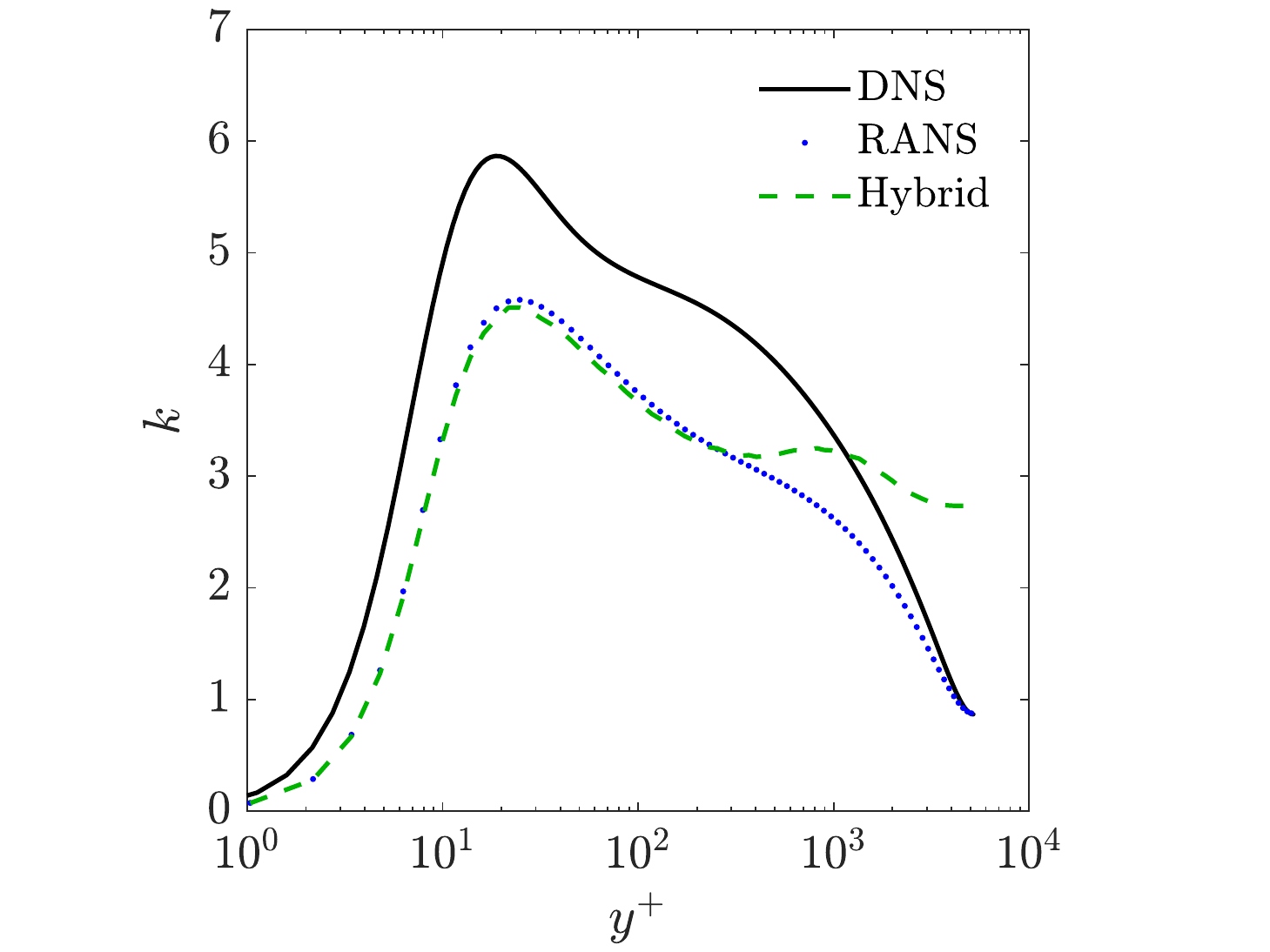}
   \caption{TKE}
\end{subfigure}
\begin{subfigure}[b]{0.49\linewidth}
\includegraphics[trim={1.25cm 0cm 1.5cm 0cm},clip=true,width=0.99\linewidth]{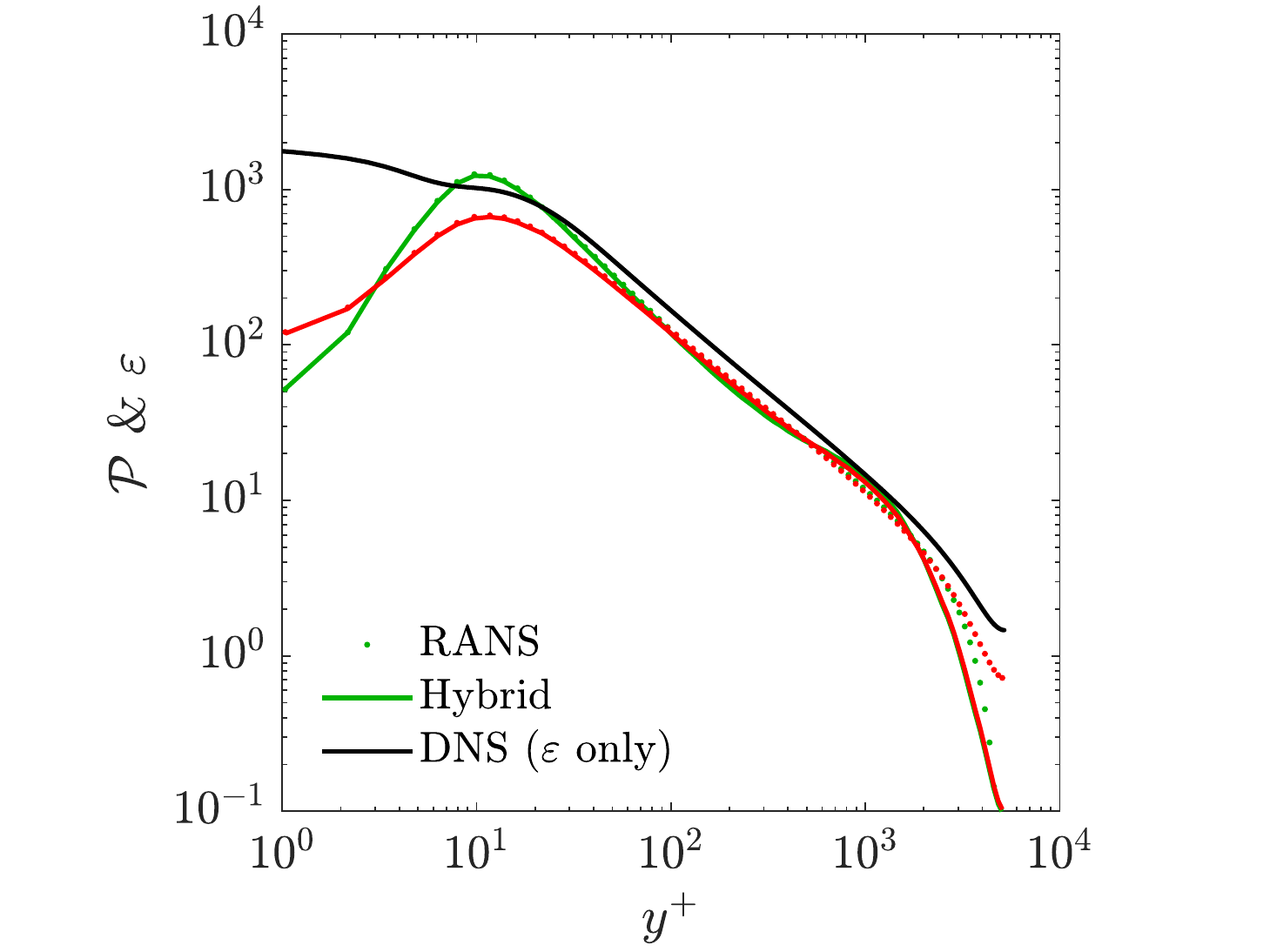}
\caption{$k$ source terms}
\end{subfigure}
\begin{subfigure}[b]{0.49\linewidth}
\includegraphics[trim={1.25cm 0cm 1.5cm 0cm},clip=true,width=0.99\linewidth]{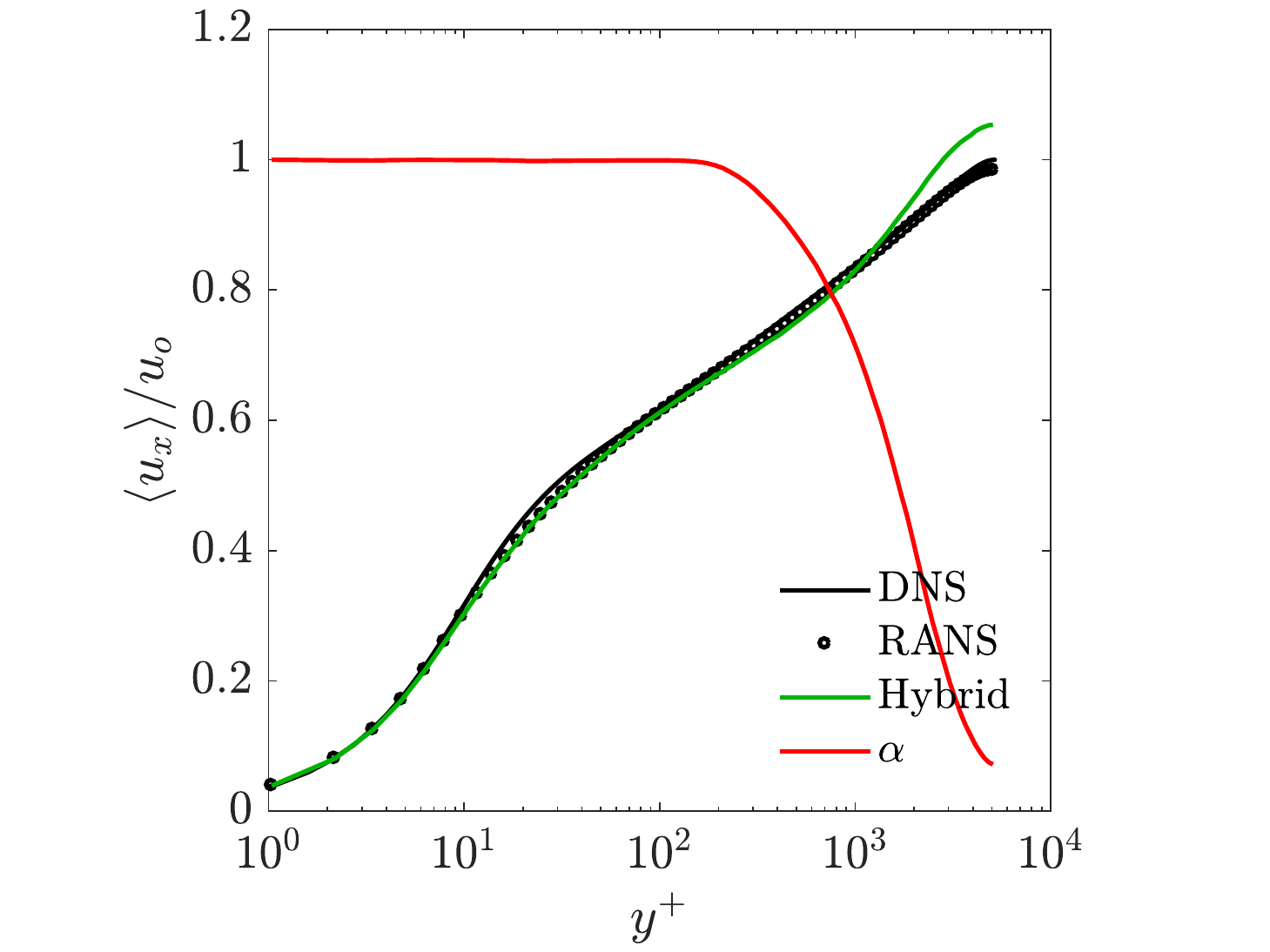}
\caption{Mean velocity}
\end{subfigure}
\end{center}
\caption{ Turbulent kinetic energy (a), production
  and destruction of TKE (b), and mean velocity (c) for turbulent
  channel flow.  For (b), the black solid line is the DNS dissipation
  and (\textcolor{green}{-}) is used for production while
  (\textcolor{red}{-}) is used for dissipation.}
\label{fig:chan_ke}
\end{figure}

Results from hybrid simulations of channel flow at $Re\approx{}5200$ (see Appendix, ``Model C'' for model details) confirm the dramatic 
effect of fluctuations on the mean source terms in the $k$ and $\varepsilon$ equations.  Figure~\ref{fig:chan_expected_sources} shows the difference between
the
mean of the fluctuating production of
$k$ and $\varepsilon$, and destruction of $\varepsilon$ and the 
values of these quantities evaluated at the mean.
As the fraction of turbulent kinetic energy that is modeled ($\alpha$)
goes down, $\mathcal{P}_k$ and $\mathcal{D}_\varepsilon$ are
enhanced by the fluctuations, and $\mathcal{P}_\varepsilon$ is
diminished. In a channel flow, the result is that in the outer region,
where $\alpha$ is small, $k$ is enhanced and $\varepsilon$
reduced relative to RANS. This modifies the subgrid stress, which then
distorts the mean velocity, as shown in Figure~\ref{fig:chan_ke}.

Transport-based 
$k$-SGS models have been used for some time \cite{dear:1973}, without
this difficulty. However, in those cases, the model length scale is
taken as the grid scale, which results in production and dissipation of
kinetic energy scaling like $k^{1/2}$ and $k^{3/2}$, respectively. The
stronger dependence of dissipation on $k$ effectively damps the $k$
fluctuations, minimizing the effects of this problem. But in a hybrid context, this
would cause new problems, such as resolved and modeled turbulence that
is out of equilibrium due to MSD, and the lack of an intrinsic model
length scale to use in hybridization. For instance, when the SGS lengthscale is 
between the integral and grid lengthscales, as will occur in hybrid simulations, 
the production scaling with $k^{1/2}$ is entirely incorrect. To fully address this problem,
the RANS transport equations should be used to represent mean
quantities as a function of only mean quantities, as designed.

\subsubsection{Stress/dissipation inconsistency in subgrid-viscosity-based LES}
In addition to problems that stem from using RANS-like transport
equations with resolved fluctuations, there are also problems that
arise when typical subgrid viscosity-based models are used in LES
regions where a significant fraction of the mean turbulent stress is
not resolved.

The issue here is an inherent inconsistency between the requirements
of predicting the subgrid contribution to the Reynolds stress and the
dissipation (or, more correctly, the transfer of kinetic energy from resolved to
unresolved scales) when using a subgrid-viscosity-based LES model, as
discussed by Jim{\'e}nez and Moser~\cite{Jimenez2000}.  In particular,
because of the model structure, the subgrid stress and strain are
fully correlated, which is physically incorrect.  Analysis suggests a
correlation of $20-30\%$ for isotropic turbulence.  This incorrect correlation would lead to
incorrect dissipation except that the models have calibration
coefficients, e.g., $C_S$ for Smagorinsky, whose value can be selected
to give the correct dissipation.  However, this calibration then
causes the subgrid Reynolds stresses to be incorrect---i.e., it is not possible
to simultaneously match both dissipation and subgrid Reynolds stresses in
eddy-viscosity-based models.  The outcome is that the subgrid contributions
to the Reynolds stress
predicted by typical LES models are often extremely wrong.  For
instance, Jim{\'e}nez and Moser~\cite{Jimenez2000} show that dynamic
Smagorinsky predicts a subgrid Reynolds shear stress that is roughly a factor
of five lower than shear stresses obtained from filtered DNS for
channel flow.

In pure LES, the impact of this problem can be minimized by using
sufficient resolution so that the subgrid shear stresses are
negligible.  In this case, only the dissipation is important.
Alternatively, in pure RANS these issues do not arise because the 
model is only required to predict the Reynolds stress and not the 
dissipation.  Unfortunately, the problem is unavoidable in hybrid 
methods.  In regions with a modest level of resolved turbulence, but not 
enough to dominate the total stress, the model must accurately 
represent both the energy transfer from resolved to modeled
scales and the subgrid contribution to the total stress. These 
observations suggest that blending through a single
eddy-viscosity-based stress model does
not provide a sufficiently rich modeling structure to accurately model
both roles of the subgrid term.

\subsection{Active Turbulence Management} \label{sec:forcing_motivation}
While a new paradigm for combining RANS and LES is necessary, this
alone is not sufficient for a predictive hybrid model. Recent hybrid
simulations of the NASA wall-mounted hump (WMH) test
case~\cite{green:2006} indicate that a mechanism for active exchange of
energy between the resolved and modeled scales is also necessary.
This conclusion is based on results from two variants of a hybrid
model (see Appendix for model details) developed by the authors.  Both of these models accomplish blending
by lowering the model stress on the time scale
associated with the modeled turbulence in regions where turbulence can
be supported.  The idea being that by lowering the stress gradually the
natural instabilities of the flow would produce resolved fluctuations
to avoid stress depletion.

\begin{figure}[htp!]
\begin{center}
\begin{subfigure}{0.49\linewidth}
\includegraphics[trim={0.2cm 9.25cm 0.2cm 10cm},clip=true,width=0.95\linewidth]{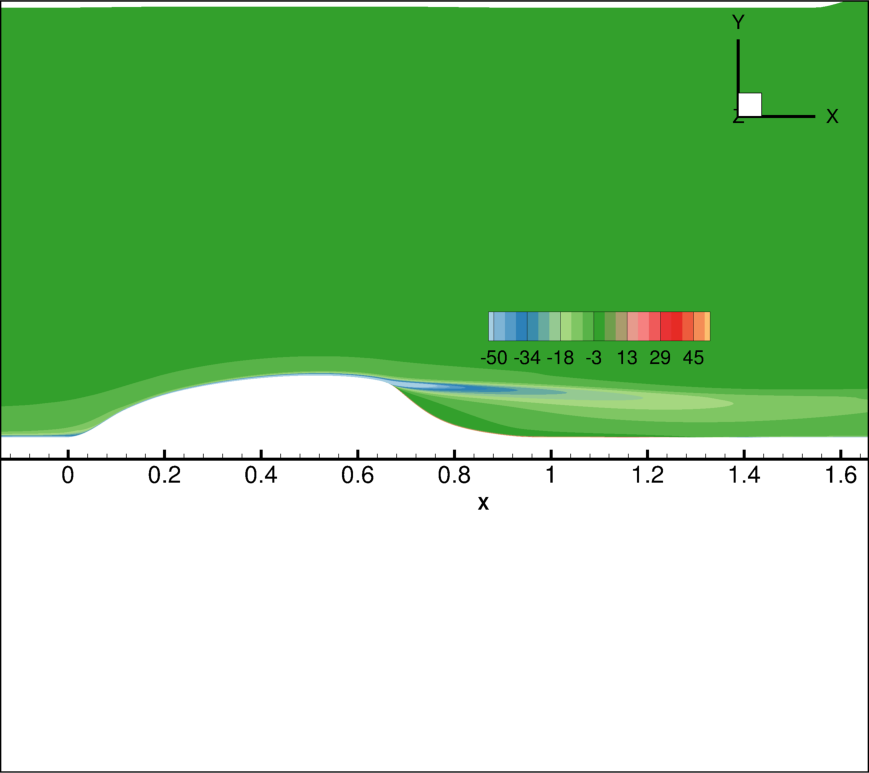}
\put(-215,45){\textcolor{white}{\small{(a) RANS $\omega_z$}}}
\end{subfigure}
\begin{subfigure}{0.49\linewidth}
\begin{overpic}
[trim={0.2cm 9.25cm 0.2cm 10cm},clip=true,width=0.95\linewidth]{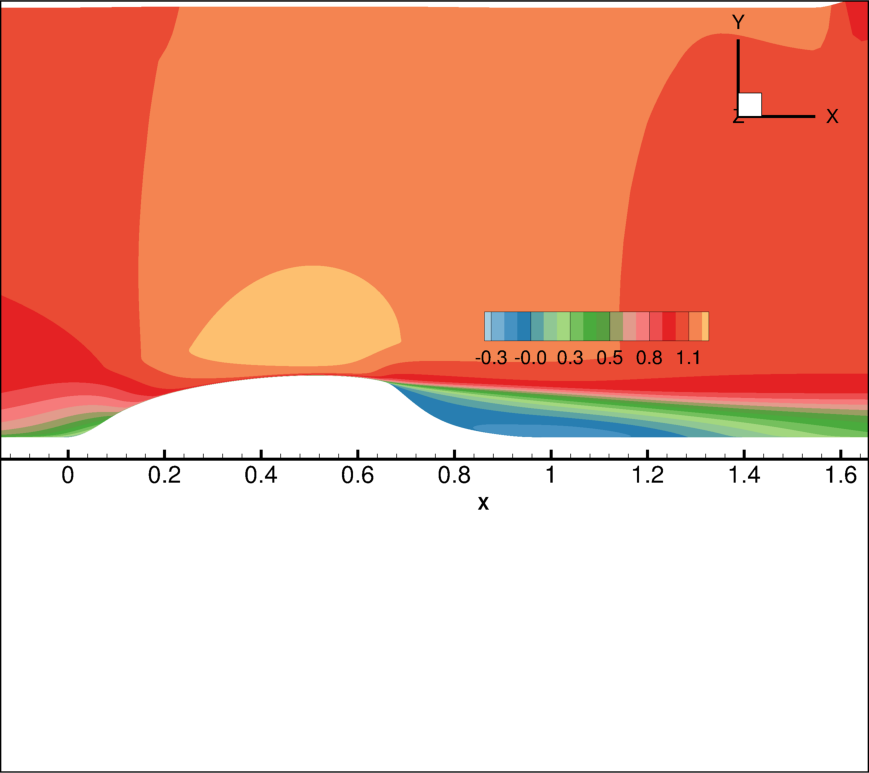}
\put(79,14){\color{black}\vector(0,-1){5}}
\end{overpic}
\put(-215,45){\textcolor{black}{\small{(b) RANS $\la{}u_x\ra$}}}
\end{subfigure}
\begin{subfigure}{0.49\linewidth}
\includegraphics[trim={0.2cm 9.25cm 0.2cm 10cm},clip=true,width=0.95\linewidth]{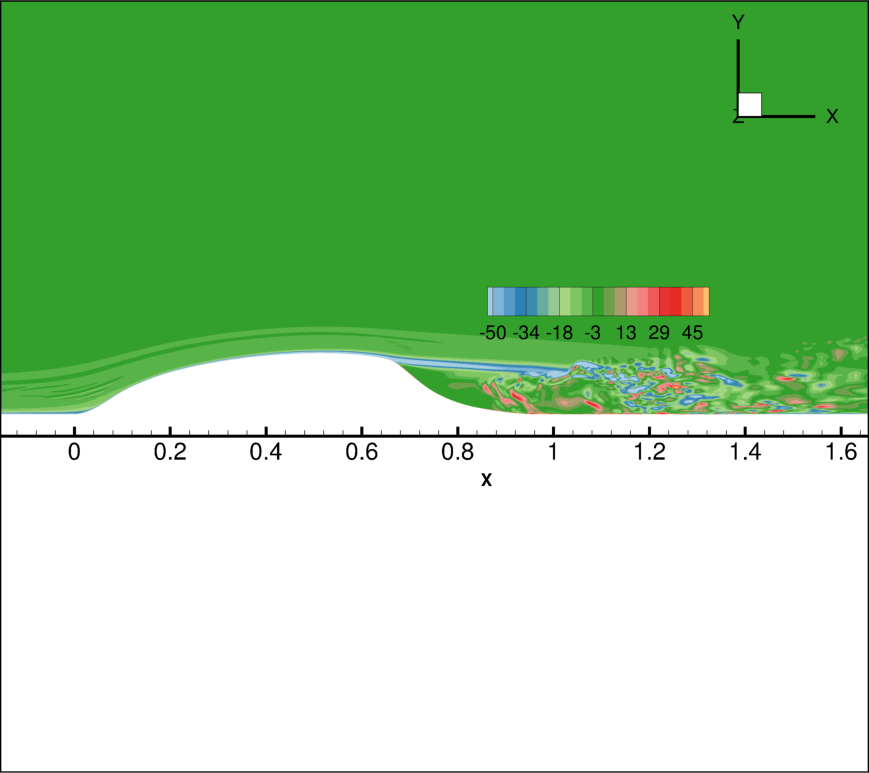}
\put(-215,45){\textcolor{white}{\small{(c) DDES $\omega_z$}}}
\end{subfigure}
\begin{subfigure}{0.49\linewidth}
\begin{overpic}
[trim={0.2cm 9.25cm 0.2cm 10cm},clip=true,width=0.95\linewidth]{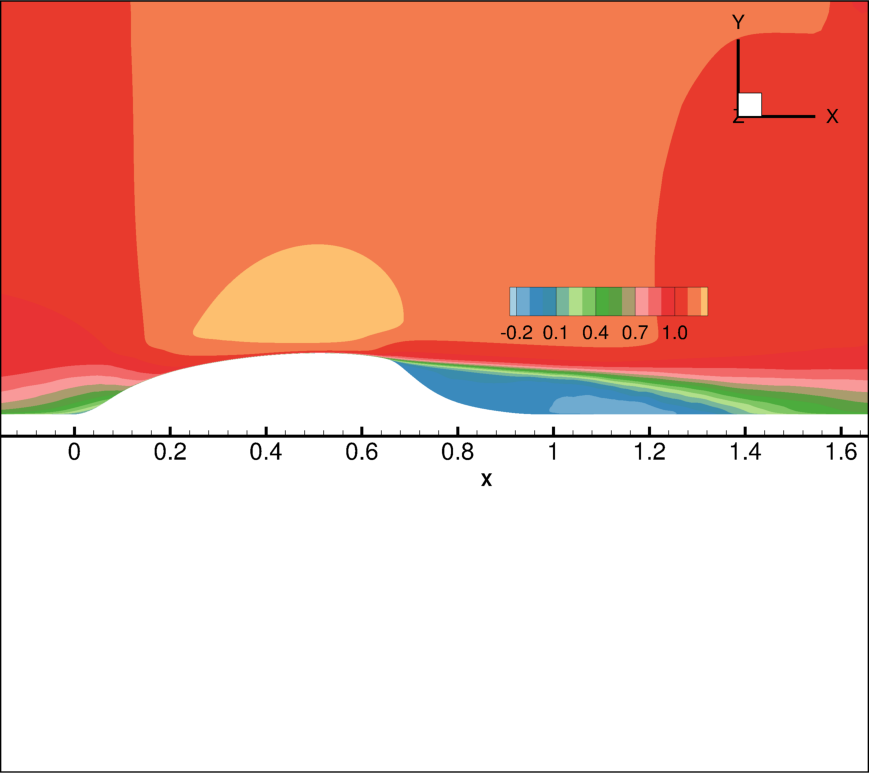}
\put(85,16.5){\color{black}\vector(0,-1){5}}
\end{overpic}
\put(-215,45){\textcolor{black}{\small{(d) DDES $\la{}u_x\ra$}}}
\end{subfigure}
\begin{subfigure}{0.49\linewidth}
\includegraphics[trim={0.2cm 9.25cm 0.2cm 10cm},clip=true,width=0.95\linewidth]{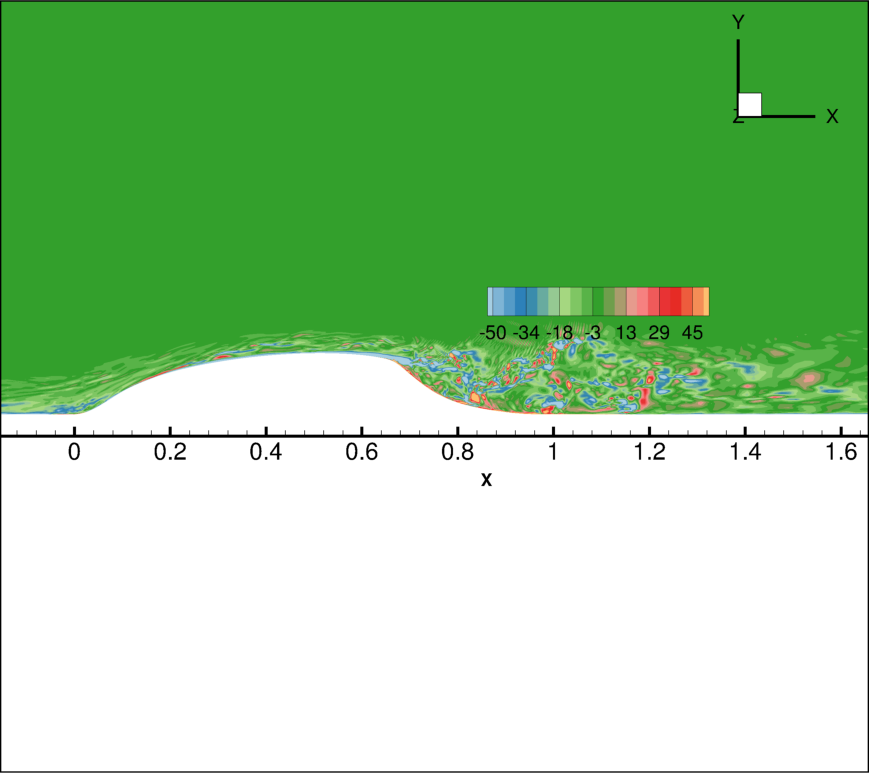}
\put(-215,45){\textcolor{white}{\small{(e) Model A $\omega_z$}}}
\end{subfigure}
\begin{subfigure}{0.49\linewidth}
\begin{overpic}
[trim={0.2cm 9.25cm 0.2cm 10cm},clip=true,width=0.95\linewidth]{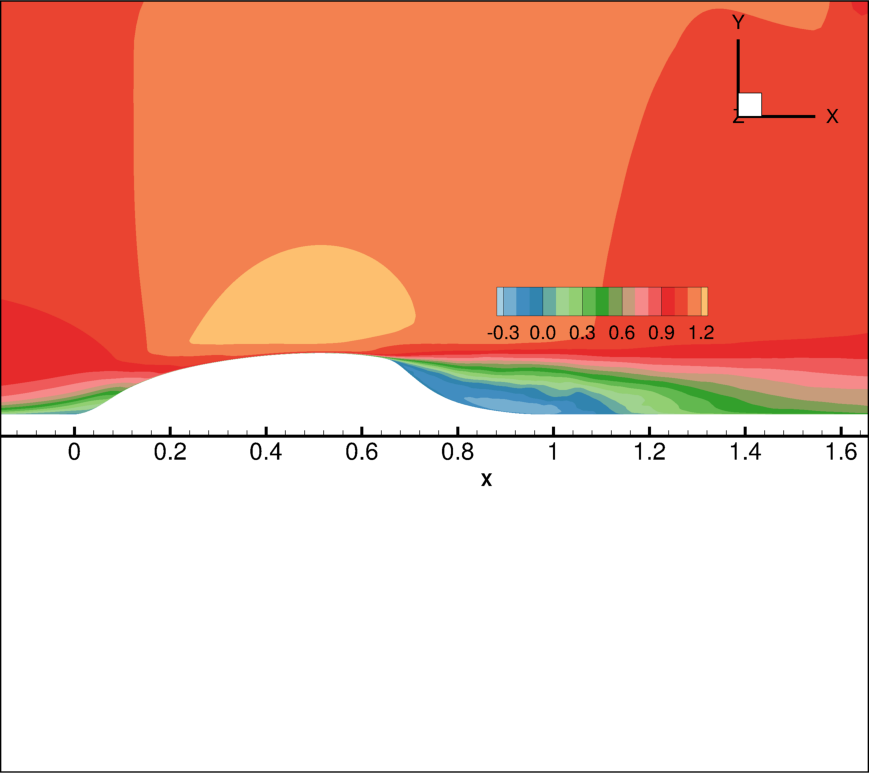}
\put(71,16.5){\color{black}\vector(0,-1){5}}
\end{overpic}
\put(-215,45){\textcolor{black}{\small{(f) Model A $\la{}u_x\ra$}}}
\end{subfigure}
\begin{subfigure}{0.49\linewidth}
\includegraphics[trim={0.2cm 9.25cm 0.2cm 10cm},clip=true,width=0.95\linewidth]{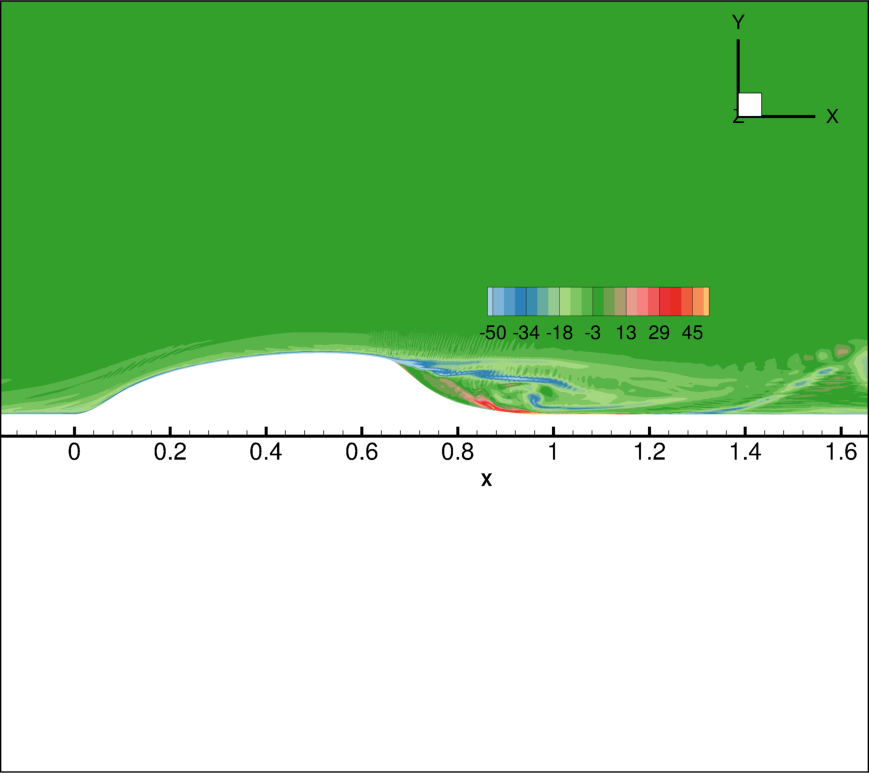}
\put(-215,45){\textcolor{white}{\small{(g) Model B $\omega_z$}}}
\end{subfigure}
\begin{subfigure}{0.49\linewidth}
\begin{overpic}
[trim={0.2cm 9.25cm 0.2cm 10cm},clip=true,width=0.95\linewidth]{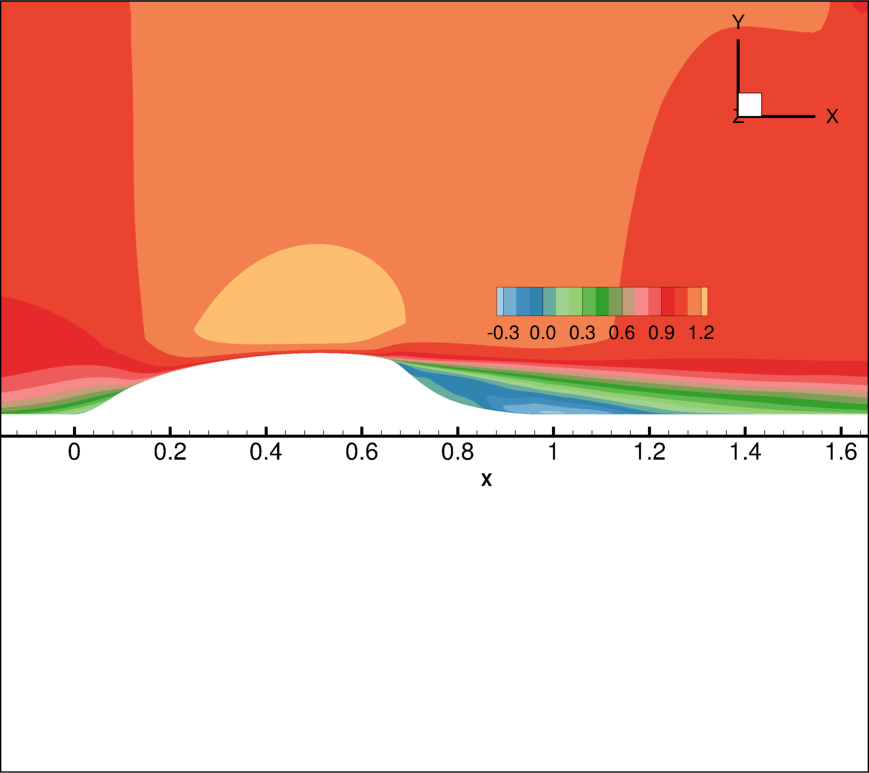}
\put(75,16.5){\color{black}\vector(0,-1){5}}
\end{overpic}
\put(-215,45){\textcolor{black}{\small{(h) Model B $\la{}u_x\ra$}}}
\end{subfigure}
\begin{subfigure}{0.49\linewidth}
\includegraphics[width=0.95\linewidth]{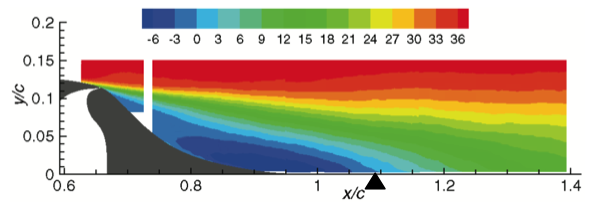}
\put(-45,17){\textcolor{white}{\small{(i) PIV}}}
\end{subfigure}
\end{center}
\caption{Results for two hybrid model variants used to simulate the NASA
  wall-mounted hump, compared to $\overline{v^2}$-$f$ RANS, DDES, and
  experiment: Instantaneous spanwise vorticity (a,c,e,g) and mean
  streamwise velocity (b,d,f,h,i).  The black arrows indicate
  reattachment location.}
\label{fig:hump_flow}
\end{figure}
\begin{figure}[htp!]
\begin{center}
\begin{subfigure}{0.49\linewidth}
\begin{center}
\includegraphics[trim={2cm 0cm 2.5cm 0cm},clip=true,width=0.77\linewidth]{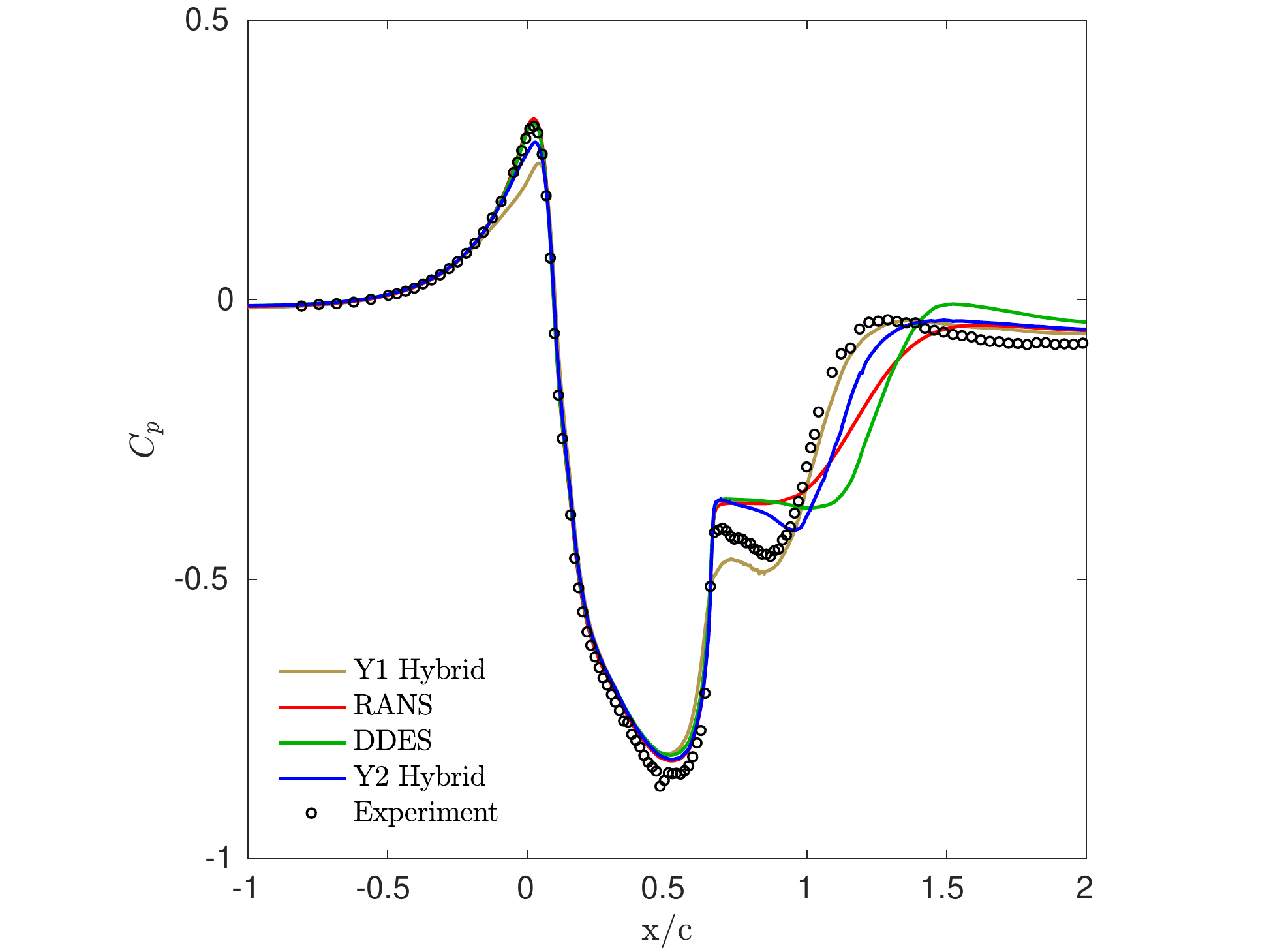}
\caption{$C_p$}
\end{center}
\end{subfigure}
\begin{subfigure}{0.49\linewidth}
\begin{center}
\includegraphics[trim={2cm 0cm 2.5cm 0cm},clip=true,width=0.8\linewidth]{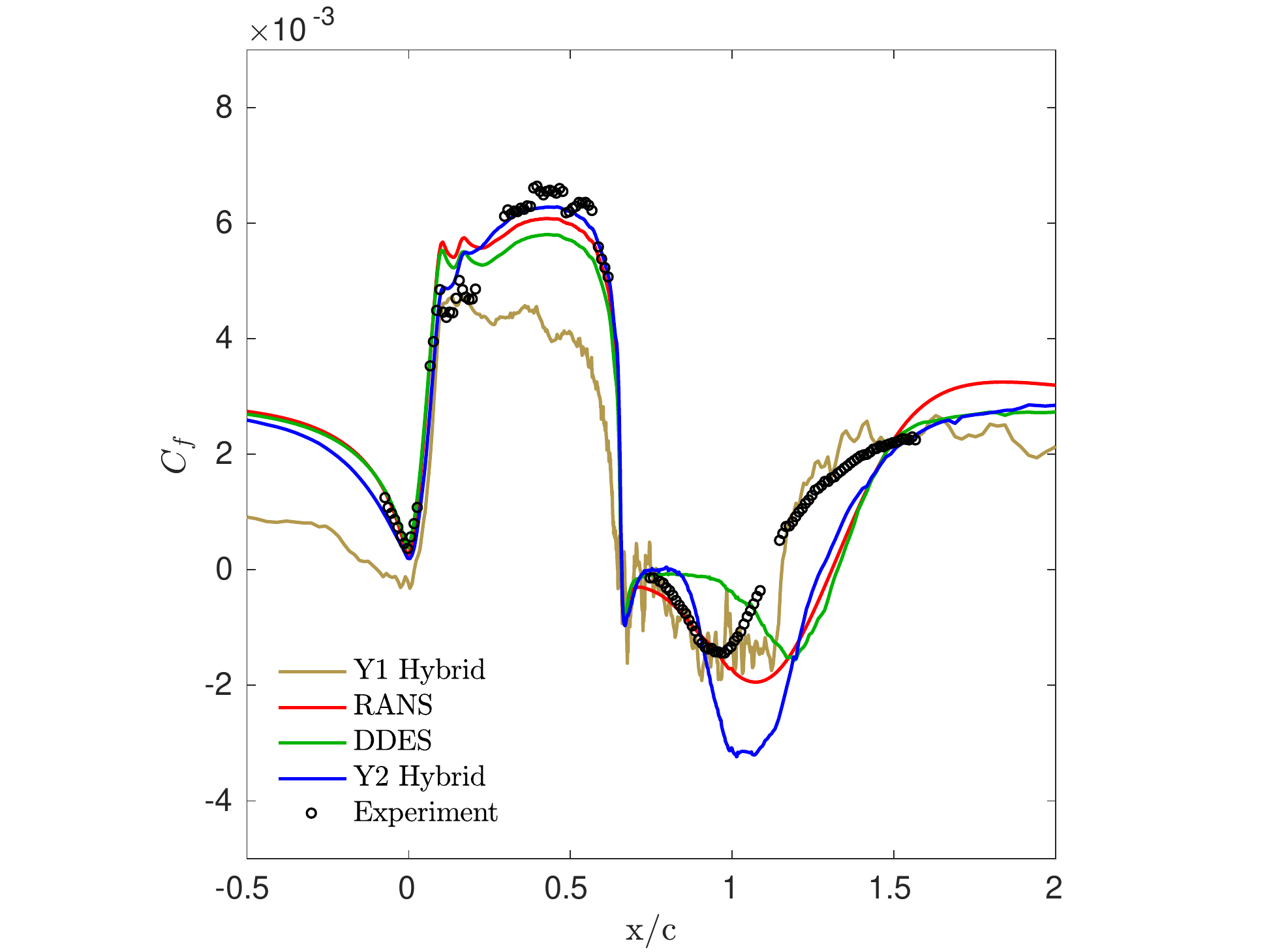}
\caption{$C_f$}
\end{center}
\end{subfigure}
\end{center}
\caption{Model A (Y1) and Model B (Y2) NASA wall-mounted hump results, compared to
  $\overline{v^2}$-$f$ RANS and DDES: pressure coefficient (left)
  and skin friction coefficient (right).}
\label{fig:hump_cpcf}
\end{figure}

Results are shown in Figure~\ref{fig:hump_flow} and~\ref{fig:hump_cpcf}.  In
the first version of the model, labeled Model A, some
resolved fluctuations are obtained in the outer portion of the
boundary layer on the hump upstream of the separation point.  These
fluctuations are essential to initiate the correct break-up of the separated
shear layer necessary to accurately predict the reattachment point.
The Model A prediction of the reattachment point agrees well with 
experiments.  However, Model A leads to a poor prediction of the
skin friction coefficient upstream of and on the hump 
(Fig.~\ref{fig:hump_cpcf}), indicating that the fluctuations are not 
sufficient to offset the drop in wall normal momentum transfer due to reduced modeled stress.


This stress depletion problem can be avoided by enhancing the modeled
wall-normal transport as is done via an anisotropic eddy viscosity in
Model B leading to accurate skin friction upstream of the separation.
However, this modification also suppresses fluctuations upstream of
the separation, keeping the solution more RANS-like in that region.
The resulting lack of fluctuations flowing into the separated shear
layer leads to delayed roll-up and an inaccurate prediction of the
reattachment point.  Qualitatively, this behavior is similar to DDES,
which is also RANS-like near the separation point and reattaches too
far downstream.

The takeaway from these results is that to correctly predict both skin friction and shear 
layer roll-up simultaneously, their must be resolved turbulence in the outer portion of the 
boundary layer and the model must ensure the resolved and modeled turbulence are in 
agreement; the only way to achieve this everywhere in a hybrid simulation is by actively 
creating resolved turbulence where the grid can support fluctuations, particularly in low 
shear regions where the instabilities in the flow are insufficient to rapidly self-generate 
turbulence.


\subsection{Effects of Anisotropy in LES}
It is well understood that standard SGS models formulated
using assumptions of isotropic subgrid turbulence and isotropic
resolution break down when these conditions are violated.  
In previous work~\cite{haer:2018}, a number of SGS models were
evaluated in forced, homogeneous, isotropic turbulence on anisotropic
grids.  The results show that, for mildly anisotropic cells, the
Smagorinksy model leads to under dissipation in coarse directions,
incorrect energy distribution across fine scales manifesting in an
incorrect inertial range slope, and over dissipation in the fine
directions.  Being under dissipative in coarse directions and at lower
wavenumbers in the fine directions is especially problematic as it
leads to spurious oscillations and can corrupt an entire simulation.
These problems are particularly acute in hybrid simulations of complex
geometries because highly anisotropic, coarse resolution is typically
present.

Elimination of excess energy at lower wavenumbers is the most critical
component in selecting the form of the SGS model anisotropy.  More
recent subgrid models, such as the anisotropic minimum dissipation
(AMD) model~\cite{roze:2015} and the M43 model~\cite{haer:2018} have,
to some extent, overcome the problems observed with Smagorinsky for
isotropic turbulence and anisotropic resolution.  However, in
traditional hybrid schemes, it is not clear how to transition between
the anisotropic behavior of these models, which are related to energy
transfer and not stress, with RANS.  This further motivates the
development of a new hybridization approach that can naturally make
use of such models.

\section{Developments Enabling Predictive Hybrid RANS/LES}
\label{sec:modeling}
This section describes a modeling approach designed to overcome the
deficiencies described in~\S\ref{sec:background}.  In particular, a
novel hybridization paradigm based on separating the RANS and standard
subgrid contributions to the model term is introduced
in \S\ref{sec:blending_proposal}.  The model requires an active energy
transfer from modeled to resolved scales.  A preliminary approach for
this forcing is outlined in~\S\ref{sec:forcing_proposal}.  Finally, an
LES model designed for use with anisotropic grids is described
in~\S\ref{sec:les_proposal}.

\subsection{Hybridization via Model-Splitting}
\label{sec:blending_proposal}

Motivated by the drawbacks of blending RANS and LES through a single,
eddy-viscosity-based stress model, as described
in~\S\ref{sec:blending_problems}, the hybridization strategy
introduced here is based upon splitting the total model stress into two
separate portions: one primarily responsible for the mean subgrid
stress and one mainly representing energy transfer to the subgrid.
Thus, in this ``model-split'' (MS) hybridization, the resolved momentum
equation takes the following form:
\begin{equation*}
D_t\overbar{u}_i
= 
-\frac{1}{\rho}\partial_i{}\bar{p}
+
\nu\partial_j\partial_j\overbar{u}_i
+
\partial_j \left( \tau^{SGRS}_{ij} + \tau^{SGET}_{ij} \right) +F_i,
\label{ms}
\end{equation*}
where $\overbar{u}_i$ is the resolved velocity, $\overbar{p}$ is the
resolved pressure, $\tau_{ij}^{SGRS}$ is intended to model the mean subgrid
stress, $\tau_{ij}^{SGET}$ is used to represent the energy transfer
from the resolved to modeled scales, and $F_i$ is used to induce the transfer
of energy from the modeled to resolved scales.   
To further develop this approach, the total velocity field is
decomposed as $u_i=\la{}u\ra_i+u'_i=\la{}u\ra_i+u^>_i+u^<_i$, where
$\la{}u\ra_i$ is the Reynolds averaged velocity, and turbulent
velocity fluctuations ($u'_i$) are divided into resolved ($u^>_i$) and
discarded ($u^<_i$) components; \emph{i.e.}, the total resolved velocity field
is $\bar{u}_i = \la{}u\ra_i+u^>_i$.  The resolved momentum flux is
$(u^>_i\la{}u\ra_j+\la{}u\ra_iu^>_j+u^>_iu^>_j)$ so that the average turbulent stress to be
modeled is $\tau_{ij}^{SGRS}\approx\la{}u'_iu'_j\ra-\la{}u^>_iu^>_j\ra$.
Assuming that both stress tensors scale with their trace,
then $\tau_{ij}^{SGRS}\approx\alpha\la{}u'_iu'_j\ra$ where $\alpha$ is
the ratio of modeled to total TKE:
$\alpha=k_{sgs}/k_{tot}=(1-k_{res}/k_{tot})$, where $k_{res}=\la{}u^>_iu^>_i\ra/2$.
Hence, a simple model for the average subgrid stress would be
$\tau_{ij}^{SGRS}=\alpha\tau_{ij}^{RANS}$, where $\tau_{ij}^{RANS}$ denotes a RANS
closure evaluated, as in standard RANS, using only mean quantities.  

This scaling represents a dramatic simplification of
the true behavior of the subgrid stress as a function of the Reynolds
stress and resolved turbulence.  The true scaling likely depends on proximity
of the subgrid lengthscale to the integral and Kolmogorov lengthscale.  
Indeed, the similarity analysis of
Lumley \cite{luml:1967}, as discussed in \cite{Jimenez2000}, suggests
that, for even low mean shear, the anisotropic portion of the modeled
shear stress decays like $\alpha^2$, i.e., faster than the
subgrid $k$.  Thus, the model is known to be wrong in some sense.
However, when coupled with existing RANS models, scaling the Reynolds
stress anisotropy by $\alpha^2$ was found to perform worse than
$\alpha$.  This behavior is attributed to the fact that the scalar
quantities, e.g., $k$ and $\epsilon$, in RANS models are only loosely
representative of the corresponding true quantities. RANS models
function by a series of error-canceling and coefficient tuning to
arrive at nearly the correct mean stress provided the mean
gradients; i.e., they get the ``correct'' eddy viscosity and not
the correct individual scalar contributions to the eddy viscosity.
Thus, it is possible that the theoretically correct scaling for any
given flow and level of resolved turbulence will not perform well in a
hybrid simulation built upon existing inadequate models.  Because of
this, the $\alpha$ scaling is used here, but improving, or removing
the necessity for, this scaling will be the subject of future works.

Regarding energy transfer, since $\tau_{ij}^{SGRS}$ is not fluctuating, it transfers kinetic 
energy from the mean to the unresolved scales, but not from the resolved fluctuations
to the unresolved scales.  Instead, the latter transfer is modeled through $\tau_{ij}^{SGET}$.  
Traditional LES SGS models primarily function to model this transfer with the majority 
of the turbulence stress being resolved.   Accordingly, one may use typical LES SGS 
model forms for $\tau_{ij}^{SGET}$.  In theory, any standard SGS model may be applicable 
however, we have performed MS simulation with the model in \S\ref{sec:les_proposal} here.  Energy transfer 
to the subgrid is largely a result of the interactions of scales of motion near the cutoff 
\cite{les:1996} and only indirectly depends on mean gradients. Thus, only the fluctuating 
part of the resolved velocity gradient is used to evaluate $\tau_{ij}^{SGET}$.  
For typical model forms, $\tau_{ij}^{SGET}$ 
may contribute to the mean stress due to non-vanishing correlations 
between the model viscosity and fluctuating velocity gradient tensor.
However, for some models, such as that described in~\S\ref{sec:les_proposal},
$\la \tau_{ij}^{SGET} \ra = 0$, perfectly delineating the roles of the two terms.  
If the $\tau_{ij}^{SGET}$ model is selected so that the SGET eddy viscosity is uncorrelated 
with the fluctuating gradient, the MS approach can be considered as a method 
of segregating the mean and fluctuating portions of the modeled stress and isolating 
their eddy-viscosity dependencies.  That is, the total 
model is decomposed into the mean deviatoric stress ($\tau_{ij}^{SGRS}=\la{}\tau-\frac{2}{3}k_{sgs}{I}\ra$) 
and the fluctuating portion ($\tau_{ij}^{SGET}=\tau - \tau_{ij}^{SGRS}$).



Due to the necessity of knowing $k_{tot}$, any two-equation RANS model
should be functional in the MS framework.  In a $k$-based model, the
model equations can be used unchanged except for the production term,
which is modified as follows:
\begin{equation}
\mathcal{P}_k=\la{}S_{ij}\ra\big{(}\tau^{SGRS}_{ij} + \la{}\tau^{SGET}_{ij}\ra- \la{}u^>_iu^>_j\ra\big{)},
\label{Pk}
\end{equation}
where the first term is already an expected quantity by construction
and the second term will generally be small, or zero, depending on the
selection of the SGET model (see \S\ref{sec:les_proposal}).
Naturally, all time and length scales used in the RANS model must
remain the mean values, and not subgrid.  Other modifications to the
RANS model transport equations, e.g., to use information from the
fluctuating state in the transport closures, can be imagined, but are
not used here.  To use a one equation model, such as
SA~\cite{spal:1994}, further modeling to obtain $k_{tot}$ is
necessary.  For example, one can approximate $k_{tot}$ from the eddy
viscosity and a timescale extracted directly from the mean velocity
gradient tensor.  Such an approach is likely to work in flows with one
primary inhomogeneous direction, such as channel flow or 2-D boundary
layers.  How to extend the approach to more general flows is left for
future work.

The model-split hybridization approach has numerous advantages over
traditional blending methods.  First, it uses the RANS model as
designed.  The turbulence model state variables represent mean
features of the turbulence, and the governing PDEs depend only on mean
quantities.  Thus, pathological behaviors of the RANS transport models
due to fluctuating state variables are avoided.  Second, the RANS
eddy viscosity appearing in $\tau_{ij}^{RANS}$ makes no contribution to the
dissipation of resolved fluctuations, allowing turbulence at all
resolvable scales to form naturally without being overly dissipated.
Third, nearly any combination of base mean and fluctuating models can
be used.  Because of this flexibility, advanced models are easily
incorporated, which is necessary to treat the effects of
anisotropy, as discussed in~\S\ref{sec:les_proposal}.


While there are many advantages, the model-split form introduces new
challenges as well.
For one, in the above development, we have tacitly assumed knowledge
of the expected value for all required quantities, yet this information is not
automatically available in hybrid simulations.  Thus, a method for
obtaining the mean quantities is required.  For stationary flows, a
causal time average with an exponentially decaying kernel should be
sufficient, and this approach is used here. In this case, the mean of
a quantity $\phi$ evolves according to
\begin{equation}
d_t\la\phi\ra = \frac{1}{T_{ave}}\big{(}\bar{\phi}-\la{}\phi\ra{}\big{)},
\label{ave}
\end{equation}
where the time constant is the large-scale turbulent time scale, or 
$T_{ave} = k_{tot} / \varepsilon$.  In this way, the majority of the turbulent 
fluctuations, $\phi^>$,  will be removed from $\la\phi\ra$.

\subsection{Active Energy Transfer}
\label{sec:forcing_proposal}

Based on the work described in~\S\ref{sec:forcing_motivation}, it has become clear 
that active exchange of energy between resolved and unresolved scales (forcing) is 
a requirement for a robust hybrid modeling framework.  In particular, in regions that 
are able to resolve more turbulence, velocity fluctuations should be added at the 
length scale of the smallest resolved turbulence.  On the other hand, in under-resolved
regions, $\tau_{ij}^{SGET}$ will naturally move energy from the resolved to the modeled 
scales.  In general, due to grid inhomogeneity, $\tau_{ij}^{SGET}$ alone may be insufficient 
to affect this transfer fast enough.  In this initial work, the forcing scheme is designed 
only to inject fluctuations into RANS and near-RANS regions that can support more
resolved turbulence.  Modification to $\tau_{ij}^{SGET}$ or $F_i$ to rapidly remove resolved 
energy in response to rapidly coarsening resolutions will be a subject of future work.  
Active forcing of the resolved field requires the ability to: 1) identify regions where more 
turbulence can be resolved, 2) determine the rate at which resolved kinetic energy 
should be added, and 3) specify the structure of the velocity fluctuation to be created.

Identifying regions where forcing is possible and/or needed requires evaluating 
the grid's ability to resolve more or less turbulence, which is a common
requirement of hybrid models.  Typical measures compare scalar measures of 
grid spacing (e.g., cell diagonal or volume cube-root) to a scalar turbulent length 
scale.  For anisotropic grids and/or turbulence such measures are clearly 
incomplete indicators and generally insufficient.  Instead, the evaluation should 
be based on the locally least-resolved orientation to ensure adequate resolution 
in all directions. Naturally, this measure must depend on both the grid and the 
flow state.   Here, an anisotropic resolution adequacy metric is constructed from 
the ratio of the grid spacing to the length scale of the turbulence produced by 
the model:
%
\begin{equation}
  r_\mathcal{M}
  =
  C_r\max_{\forall{}e}(\mathcal{L}^{-1}_{\mathcal{P}}\cdot\mathcal{M})
  \approx{}
  \bigg{(}\frac{3}{2\overbar{v^2}}\bigg{)}^{3/2}\max(\mathcal{P}^{sgs}_{ik}\mathcal{M}_{kj}),
  \label{rd}
\end{equation}
where $C_r$ is a constant (currently set to 1) related to how many grid sizes are required to resolved a 
turbulent structure of size $\mathcal{L}_{\mathcal{P}}$, 
$\overbar{v^2}$ is the variance of the wall-normal velocity fluctuations, $\mathcal{P}^{sgs}_{ij}$ 
is the subgrid production tensor (not the the RANS production),
\begin{equation}
\mathcal{P}^{sgs}_{ij} = \tfrac{1}{2}\big{(}\tau_{ik} \partial_j\bar{u}_k + \tau_{jk} \partial_i\bar{u}_k\big{)},
\end{equation}
where the full subgrid stress, including the mean isotropic portion, is considered 
\begin{equation}
\tau_{ij}=\tau^{SGRS}_{ij}+\tau^{SGET}_{ij}+\tfrac{2}{3}\alpha{}k_{tot}\delta_{ij}
\end{equation}
and $\mathcal{M}_{ij}$ is a tensor characterizing the anisotropic resolution (square-root of symmetric part of metric tensor defining mapping from a unit cube to the physical cell~\cite{syng:1949}).  
The resolution tensor is
composed of eigenvalues describing the cell orientation and
eigenvalues being common grid dimensions
($\delta_x,\delta_y,\delta_z$) with its invariants being common
ambiguous grid measures, \emph{i.e.}  $3\delta_{ave}=\mathcal{M}_{kk}$,
$\delta^2_{diag}=(\mathcal{M}_{ij}\mathcal{M}_{ij})$,
$\delta^3_{vol}=\det(\mathcal{M})$.
The isotropic portion of $\tau$ is typically ignored in favor of a modified pressure gradient as it 
does not actually produce any turbulent kinetic energy. However, in this context it is necessary for evaluating the 
directional production of subgrid turbulence.
The quantity $\overline{v^2}$ is naturally available in 
Durbin's $\overline{v^2}$-$f$ model~\cite{durb:1995}, but can also be approximated 
using any other two-equation turbulence model as $\overline{v^2} \approx 5\nu_t/T$, 
where $T$ is the model turbulence time scale (for instance $T\approx1/(\beta^*\omega)$ 
for the SST model).  Thus, its presence in~\eqref{rd} does not restrict the method to 
the $\overline{v^2}$-$f$ model but is merely a vehicle for incorporating near-wall 
effects. 




This resolution adequacy measure is the ratio of the resolution length
scale to the length scale of turbulence produced by the model.  Thus,
when $\la{}r_\mathcal{M}\ra > 1$, the simulation is under-resolved,
and vice versa when $\la{}r_\mathcal{M}\ra < 1$.  For isotropic grids
with isotropic turbulence in equilibrium, $\la{}r_{\mathcal{M}}\ra{}$
reduces to the standard DES length scale comparison.  Thus, it may be
viewed as a tensor generalization of this approach.  In regions where
$\la{}r_{\mathcal{M}}\ra < 1$, the goal of forcing is to introduce
some resolved fluctuation at a gradual rate allowing the forced
resolved structures to evolve into actual turbulence thereby not
corrupting the mean.  These goals and dimensional consistenty suggest
that the magnitude of the forcing acceleration should be based on the
largest of the unresolved fluctuations and the subgrid timescale.
Making use of near-wall anisotropy, the target forcing is defined as
\begin{equation}
|F_{tar}| = C_F\frac{\sqrt{\alpha\overbar{v^2}}}{\alpha{}T}.
\label{Ftar}
\end{equation}
However, uniform forcing at $|F_{tar}|$, over some region where 
$\la{}r_{\mathcal{M}}\ra < 1$, is not desirable, as it would necessarily result in forcing the mean.  
Instead, in this work, we prescribe the spatial structure of the acceleration field and allow for 
$|F_{tar}|$ to only be attained where that artificial structure and the resolved turbulence 
allow.  To this end, an auxiliary field is defined based on the structure of a Taylor-Green (TG) 
vortex  with variable length scale.  
This TG field is as follows:
\begin{align}
  h_1(x,t) &= A\cos(a_1 x_1^p) \sin(a_2 x_2^p) \sin(a_3 x_3^p), \\
  h_2(x,t) &= B\sin(a_1 x_1^p) \cos(a_2 x_2^p) \sin(a_3 x_3^p), \\
  h_3(x,t) &= C\sin(a_1 x_1^p) \sin(a_2 x_2^p) \cos(a_3 x_3^p),
  \label{TG}
\end{align}
where the magnitudes are somewhat arbitrary but are selected such that
$h_ih_i\le{}1$ with $A=1$, $B=-1/3$, and $C=-2/3$.  The vortex sizes are 
set to mimic to local length scale at the cutoff, 
$\ell=min(N_LL_{sgs},\delta_{wall})$, where $L_{sgs} = (\alpha{}k_{tot})^{3/2} / \varepsilon$, $\delta_{wall}$ 
is the distance to the wall, and $N_L = 1$.  This is accomplished by letting
\begin{equation}
a_i=\pi/(D_i/\textsf{nint}(D_i/\min(\max(\ell,2\ell^\mathcal{M}_i),D_i)))
\end{equation}
where $\ell^\mathcal{M}_i=(r_ir_i)^{1/2}$ with $r_i = \mathcal{M}_{ij}e_j$ for each global coordinate 
direction $e_i$, and $D_i$ is the domain size in periodic directions.  The consideration for periodic 
domains can be set to a very high number or simply removed ($a_i=\pi/\max(\ell,2\ell^\mathcal{M}_i)$) 
in non-periodic directions.  The TG vortex coordinates, $x_i^p$, are selected to translate with 
the mean flow to provide some degree of temporal correlation in addition to the desired 
length scale
\begin{equation}
 x_i^p(x,t) = x_i + \langle u_i \rangle{}t.
\end{equation}
With the target acceleration magnitude (\ref{Ftar}) and structure (\ref{TG}) in hand, additional 
modifications are necessary to respect RANS and DNS limits while also acknowledging the 
combination of $h_i$ and existing turbulent fluctuations would generally result in no net resolved 
production.  The latter is addressed by testing the resolved production due to $h_i$ as
\begin{equation}
\mathcal{P}^{test}_F = (|F_{tar}|\Delta{}t)h_iu^>_i,
\label{Ptest}
\end{equation}
which will be used to clip (\ref{TG}).  The RANS limit is readily apparent when $\alpha=1$ while 
identification of near-DNS conditions requires an approximation of the local Kolmogorov length 
scale.  With the information provided by the RANS model, this limit is found as 
$\alpha_{kol}=C_\nu(\nu\varepsilon)^{1/2}/k_{tot}$ with $C_\nu=1$.  The effects of these three 
limiters are incorporated in a single scaling coefficient as
\begin{equation}
\eta = -|F_{tar}|\left\{
\begin{array}{rl} & \min(S_r-D_r-F_r,0) {\ }\textsf{if}{\ } \mathcal{P}^{test}_F\ge0 \\
&0{\ }\textsf{otherwise}
    \end{array} \right.
\end{equation}
where
\begin{equation}
D_r=(1+\alpha_{kol}-\alpha)\min(S_r,0),
\end{equation}
\begin{equation}
F_r=\alpha\max(S_r,0),
\end{equation}
and
\begin{equation}
S_r = \tanh(\log(\la{}r_\mathcal{M}\ra)),
\end{equation}
is a function that gradually turns on forcing based on $\la{}r_{\mathcal{M}}\ra$.  The forcing 
vector is simply
\begin{equation}
F_i = \eta{}h_i.
\label{Fi_actual}
\end{equation}
The above outlined approach is rudimentary.  While it is
clear that some forcing is required, what form it should take is not.
The problem of introducing progressively more refined turbulent
structures in an LES requires these structures to be synthetically
generated.  One way to view $F_i$ is as a model for the commutation
error of the filtering operation and the substantial derivative,
$\overbar{D u_i}/Dt - D \overbar{u}_i/Dt$, in the momentum equation
induced by time- and space-dependent filter width.  From this
perspective, the only way to avoid ``making up'' fluctuations would be
to have the full DNS solution at every step of an LES which would, of
course, obviate the LES. 

Existing forcing methods have typically addressed this issue by generating 
fluctuations only at a prescribed LES inlet \cite{Shur2014,Spalart2017}.  In addition to 
requiring specification of the distinct LES region and entire spectrum of locally resolved synthetic turbulent scales and 
intensities, the LES inlet location must be sufficiently upstream of the actual feature 
of interest so that the artificial inflow condition can ``heal'' to a realistic state.  Body forcing 
may be thought of as a method to introduce synthetic turbulence at a ``healing'' rate so that, through the 
entire forced region, the flow maintains a realistic state.  In this way, none of the simulated domain and associated 
computational cost is sacrificed allowing the entire 
simulated solution to yield useful information.  The 
method proposed here is somewhat similar to the body forcing in the SAS-F model of 
Menter \emph{et.al.} \cite{ment:2010}.  However, explicit time step dependance 
has been removed, and the structure is based on the variable length scale TG field 
and subgrid velocity scale rather than relying on generation of a wide spectrum of 
fluctuations using the Random Flow Generator method \cite{krai:1969,batt:2004}.  By restricting 
the forcing structure to be only at the scale of the smallest locally resolved turbulence, existing 
larger, and presumably realistic, structures are not corrupted.

Nonetheless, the particular method outlined here has a number of
potential drawbacks.  First and foremost, the resulting $F_i$ is not
divergence free.  This is a result of TG fields only being
divergence-free when constructed using a uniform overall scaling and
vortex length scale.  In an incompressible solver, the dilatational
portion is projected out, which alters the structure of the effective
forcing.  The result is that the forcing can dissipate resolved
turbulence at some points in space and time.  However, on average, the
approach does result in net production of resolved fluctuations.  In a
compressible solver, the method may result in spurious acoustic
sources.  Second, it should be possible to force more strongly, i.e.,
increase the coefficient $C_F$ (currently set to 8), by making the
forcing more representative of some acceleration which would produce a
real state of turbulence with length scales between $L^n_{sgs}$ and
$L^{n+1}_{sgs}$ over some $\Delta{}t$ for the particular
$\overbar{u}_i$.  That is, if the forcing were more representative of
the commutation error $\overbar{D u_i}/Dt - D \overbar{u}_i/Dt$,
generation of realistic resolved content could be affected more
efficiently.  There is no reason to expect the augmented TG field to
be representative of these effects.  By improving upon this structure,
shorter lengths of LES resolution upstream to areas of interest, such
as flow separation, will be necessary.

\subsection{Anisotropy Treatments}
\label{sec:les_proposal}
In practical flows of engineering interest, the combination of high
Reynolds number with complex geometries and limited computational
resources often dictates discretization using relatively high
aspect ratio (AR) cells of varying shape and rapidly varying size in
space.  In such cases, the assumptions of isotropic unresolved
turbulence in equilibrium with the large scales and isotropic LES
filtering will often be violated.  Accordingly, issues of both
turbulence and resolution anisotropy must be addressed.

In previous work, we applied a resolution-scale
anisotropy representation to the entire LES model term by introducing a
direction-dependent eddy viscosity~\cite{HaeringPhd}:
\begin{equation}
  \tau_{ij}=\nu_{jm}\partial_m\bar{u}_i+\nu_{in}\partial_n\bar{u}_j,
  \label{tau_bar}
\end{equation}
with multiple tensor eddy viscosity models having been constructed and
evaluated~\cite{year2}.  In light of the model-splitting
method (\S\ref{sec:blending_proposal}), the generalization of this
approach requires two eddy viscosity tensors.  Consistent with the
purpose of each component of the subgrid model, model anisotropy can
be divided into two main categories: turbulent stress anisotropy and
dissipation anisotropy.  Stress anisotropy describes how momentum is
transferred differently between each velocity component in the flow.
Dissipation anisotropy, on the other hand, describes the different
rates of energy transfer to unresolvable scales resulting from each
velocity component and filter direction.  The former is primarily due
to the largest scales of turbulence while the latter depends on the
smallest resolved scales.  These effects are separated in the
model-split hybridization approach, and thus, separate anisotropy models can
also be used.

Since stress anisotropies are mostly carried by the largest of
turbulent structures, it is reasonable to expect that even small
resolved turbulent kinetic energy will result in adequate total stress
anisotropy in the model-splitting approach.  This is, in fact, one of
the main reason for hybrid modeling.  An important reason for poor
RANS performance is poor representation of Reynolds stress
anisotropy \cite{laun:1975}.  If the hybrid approach is correctly
formulated, such anisotropies should be captured in regions where
sufficient resolution is provided.  In this case, scalar $\nu_t$
models should be sufficient for $\tau_{ij}^{SGRS}$.  Thus, in this initial
work, a standard $\overbar{v^2}$-$f$-based RANS model, specifically
the code-friendly version\cite{lien:2001}, is used, but any of the
standard models could be used.  Anisotropic models for $\tau_{ij}^{SGRS}$
will be explored at a later time.

\begin{figure}
\begin{center}
\includegraphics[width=0.6\linewidth]{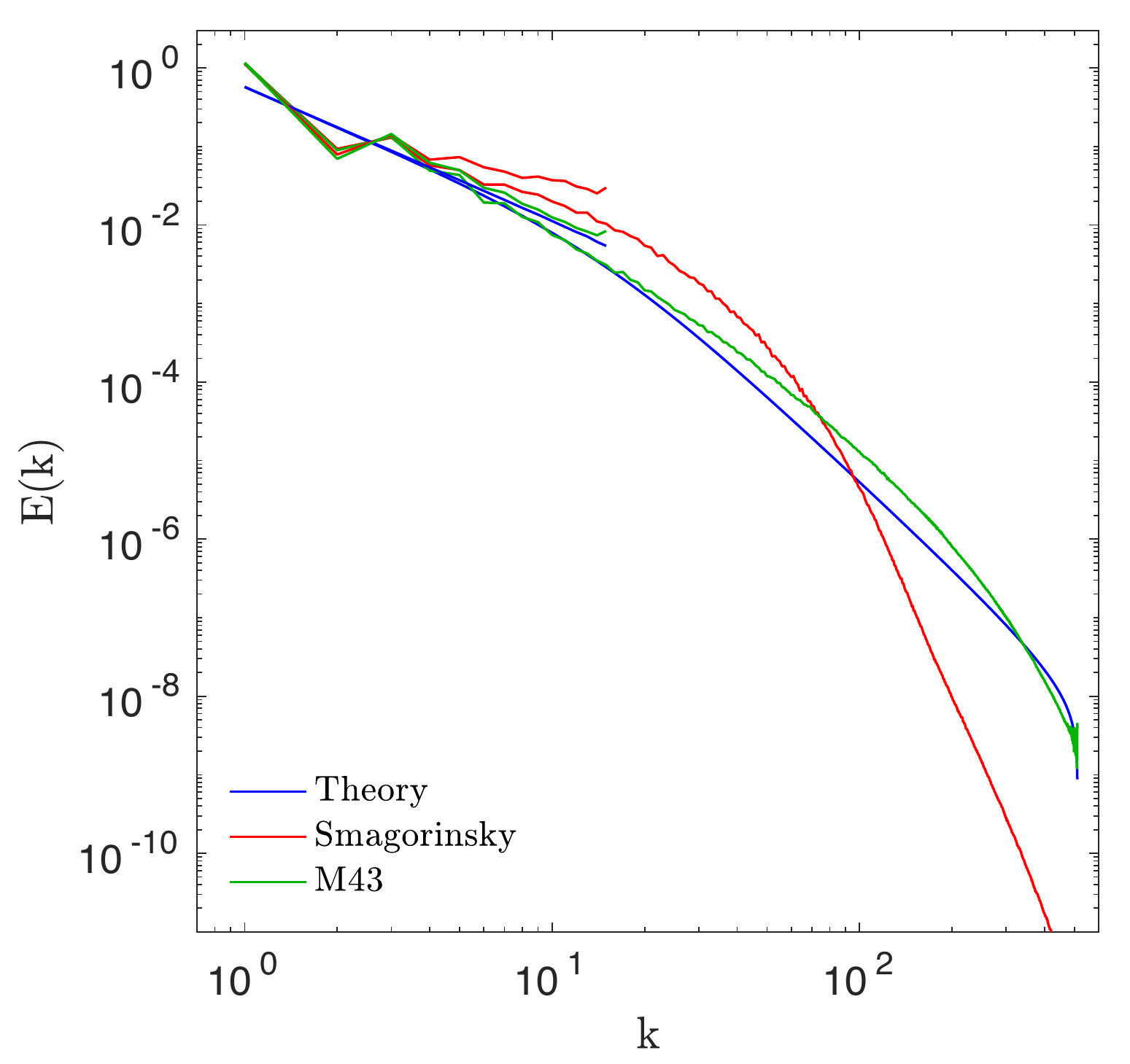}
\end{center}
\caption{Fine and coarse directional energy spectra for ($1:32:32$) AR grids and infinite $Re$ forced HIT using standard Smagorinsky and (\ref{nu_m43}).}
\label{fig:Ek}
\end{figure}

As the dissipation anisotropy is responsible for different rates of
resolved energy removal in each direction, it does directly affect the
structure of the resolved turbulence near the cutoff.  
Simulations of forced, homogeneous, isotropic
turbulence with anisotropic meshes have shown that the following eddy
viscosity tensor, labeled the ``M43'' model, gives reasonable results~\cite{haer:2018}
\begin{equation}
\nu^t_{ij}=C(\mathcal{M})\la\varepsilon\ra^{1/3}\mathcal{M}^{4/3}_{ij},
\label{nu_m43}
\end{equation}
where $\la\varepsilon\ra$ was set equal to the power input by
low-wavenumber stirring and the ``constant'' is a function of the
eigenvalues of $\mathcal{M}$ found by fitting to the expected
dissipation using the expected grad-grad product,
$\la{}\partial_ku^>_i\partial_lu^>_j\ra$, assuming a $k^{-5/3}$
inertial range spectrum.  Note that the $C(\mathcal{M})$ used here corresponds to 
second-order finite volume numerics and is not the same is of that presented in 
\cite{year2} where a psuedo-spectral code was used.  The specific coefficients 
used here are given in the appendix.  When applied to varying AR cells and 
shapes, (\ref{nu_m43}) was shown to be greatly superior to Smagorinsky and nearly
equivalent to AMD \cite{roze:2015} (Fig. \ref{fig:Ek}).  This is particularly interesting
because both of these previous scalar eddy viscosity forms are expressed in 
terms of instantaneous fluctuating velocity gradients, and therefore vary greatly and 
may contribute to the mean stress as there may be non-vanishing 
correlations.  The tensor eddy viscosity~\eqref{nu_m43} however, relies on no gradient
information and does not fluctuate.  Therefore, the fluctuating gradient portion
does not contribution to the mean stress.  With the $\overbar{v^2}$-$f$ model used 
here, the required expected dissipation is set to the RANS model $\varepsilon$.

In the model-split hybridization approach, this model form can be used for
$\tau^{SGET}$ with the necessary mean dissipation provided by the RANS
model.  This choice is used in the example in~\S\ref{sec:results}.

\subsection{Similarities with Existing Models}
Though the proposed hybrid structure is unique in total with particularly novel components, it does share 
qualities with existing methods and, in the case of DES, expanded on existing concepts.  The anisotropic resolution adequacy 
parameter collapses to the lengthscale comparison driving all DES methods under 
the conditions of isotropic resolution and equilibrium/isotropic turbulence.  
The DHRL method \cite{bhus:2012} similarly divides the total modeled 
stress into RANS-based and LES-based models but retains dependance 
on the total resolved velocity field for each contribution while introducing 
blending between each term based on their resulting production.   The 
dual-mesh hybrid methods of \cite{jenn:2017} explicitly solves a RANS 
and LES set of equations on disparate grids and also uses time-averaging 
to enforce consistency between the resolved and modeled fields by 
modification of the RANS stress with the average of the resolved Reynolds 
stress present in the LES equations.  Such an approach does provide 
and improved mean but it is unclear how it used to improve the turbulence 
modeling beyond the changes to mean terms.

\section{Demonstration}
\label{sec:results}

To demonstrate the potential of the proposed method, fully-developed, incompressible, 
turbulent channel flow at $Re_\tau\approx5200$ is simulated.  DNS data is available for 
this case~\cite{lee:2015} allowing evaluation of the hybrid results.  A branch of the existing 
finite volume incompressible Navier Stokes solver CDP v2.4 \cite{ham:2004,you:2008} 
developed at the Stanford CTR is used here.  The solver is $2^{nd}$ order in time and space 
with no upwinding used for convective fluxes in the momentum equation and time advancement 
performed with Crank-Nicolson.  For the model-split hybridization, the SGRS model is based 
on Durbin's $\overbar{v^2}$-$f$ RANS model~\cite{durb:1995} with the redistribution rate 
model modified as in the appendix.  The SGET model used is the aforementioned M43 
model (\ref{nu_m43}) described in~\S\ref{sec:les_proposal}. 



The domain for the hybrid simulation is $2 \pi \times \pi$ (streamwise
$\times$ spanwise), which is a substantial truncation relative to the
DNS domain, which is $8 \pi \times 3 \pi$.  Relative to the DNS
resolution, the resolution of the hybrid simulation is reduced by a
factor of $(25, 10, 64)$ in the streamwise, wall-normal, and spanwise
directions resulting in grid spacing of  $\Delta^+_x\approx350$, $\Delta^+_z\approx400$, and $\Delta^+_y\approx300$ in the channel center with $\Delta^+_y\approx1$ at the first cell.  This combination of the domain truncation and grid
coarsening leads to a grid size reduction of nearly a factor of $2 \times 10^5$.

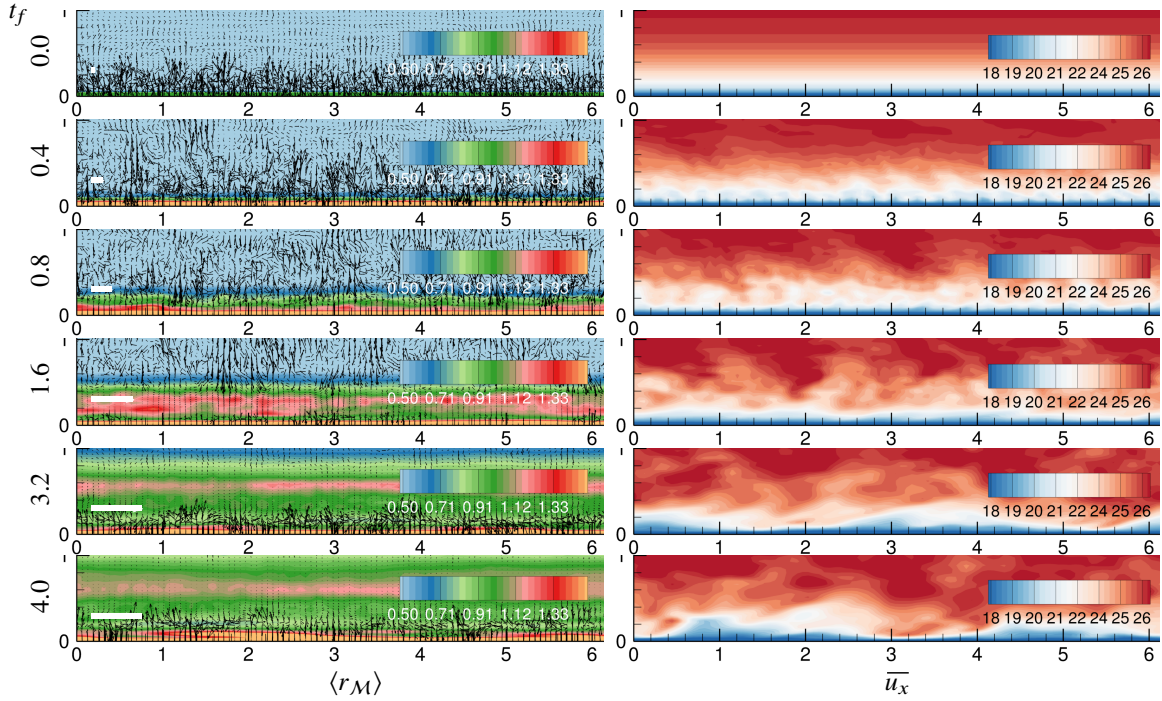
\begin{figure}[htp!]
\begin{center}
\begin{overpic}
[trim={7cm 5cm 10cm 48.5cm},clip=true,width=0.45\linewidth]{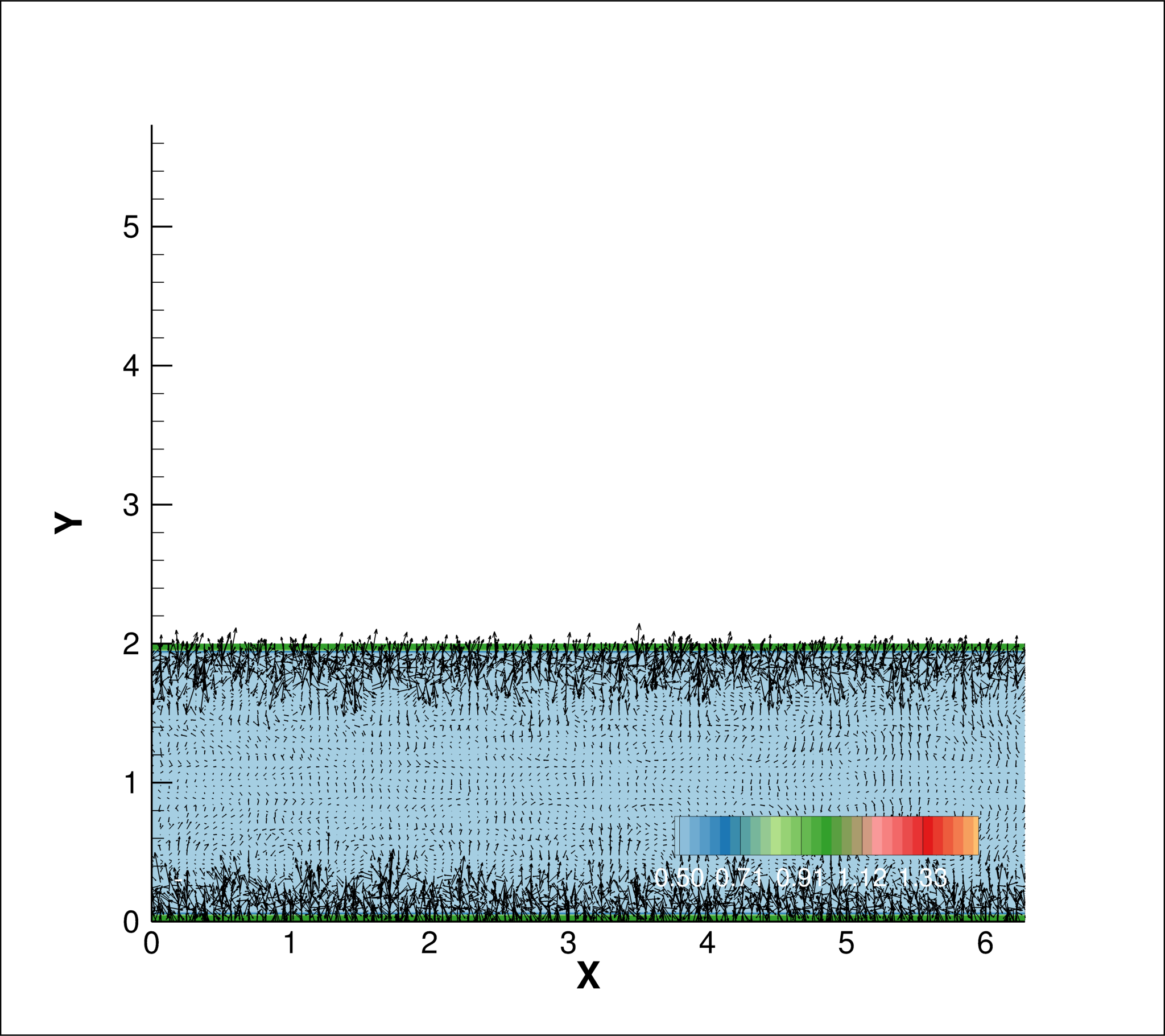}
\put(7.0,8.0){\color{white}\rule{1.3pt}{2pt}}
\put(-4.0,9.0){\color{black}\rotatebox{90}{0.0}}
\put(-8.0,18.0){\color{black}\rotatebox{0}{$t_f$}}
\end{overpic}
\includegraphics[trim={7cm 5cm 10cm 48.5cm},clip=true,width=0.45\linewidth]{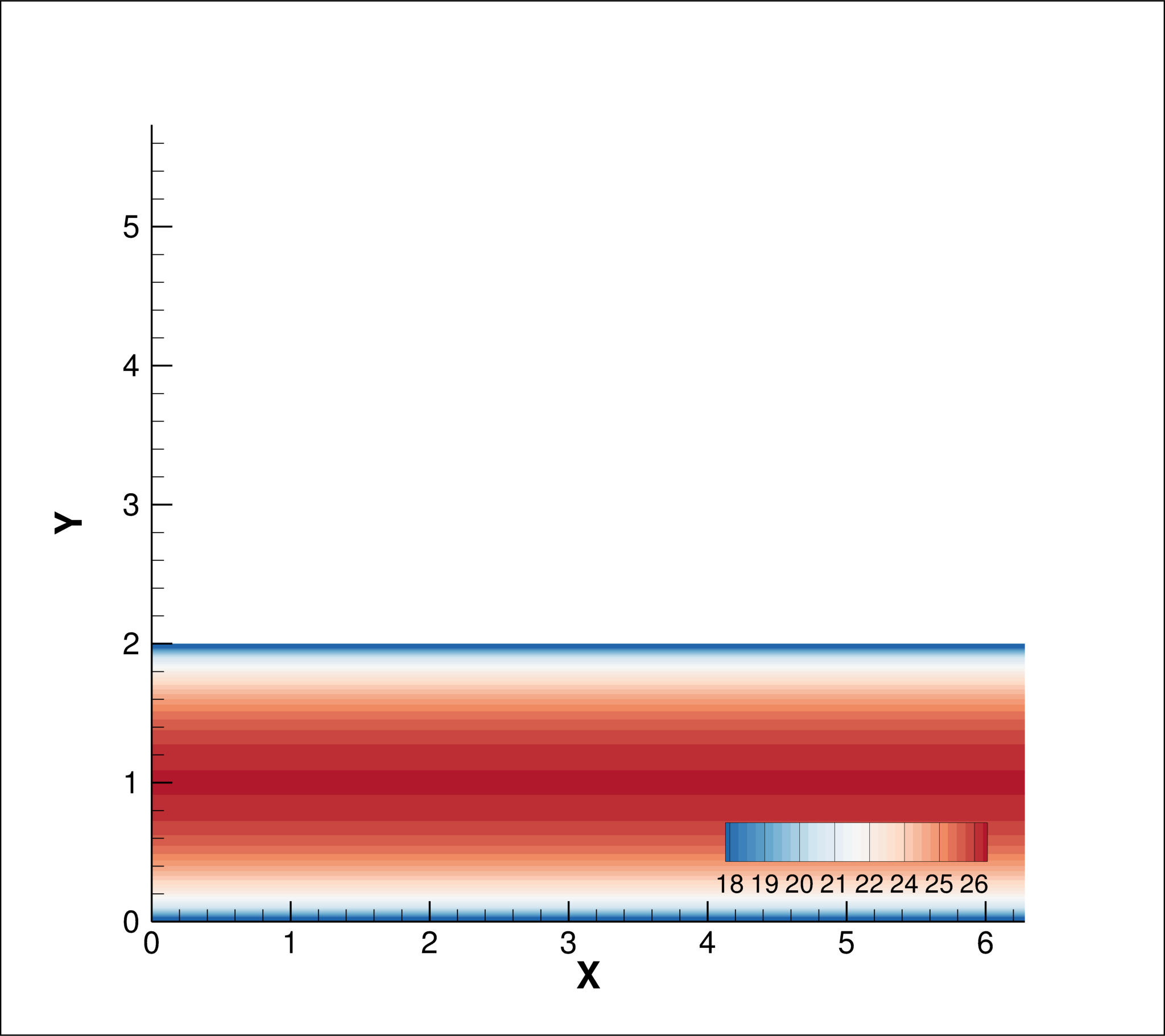}
\begin{overpic}
[trim={7cm 5cm 10cm 48.5cm},clip=true,width=0.45\linewidth]{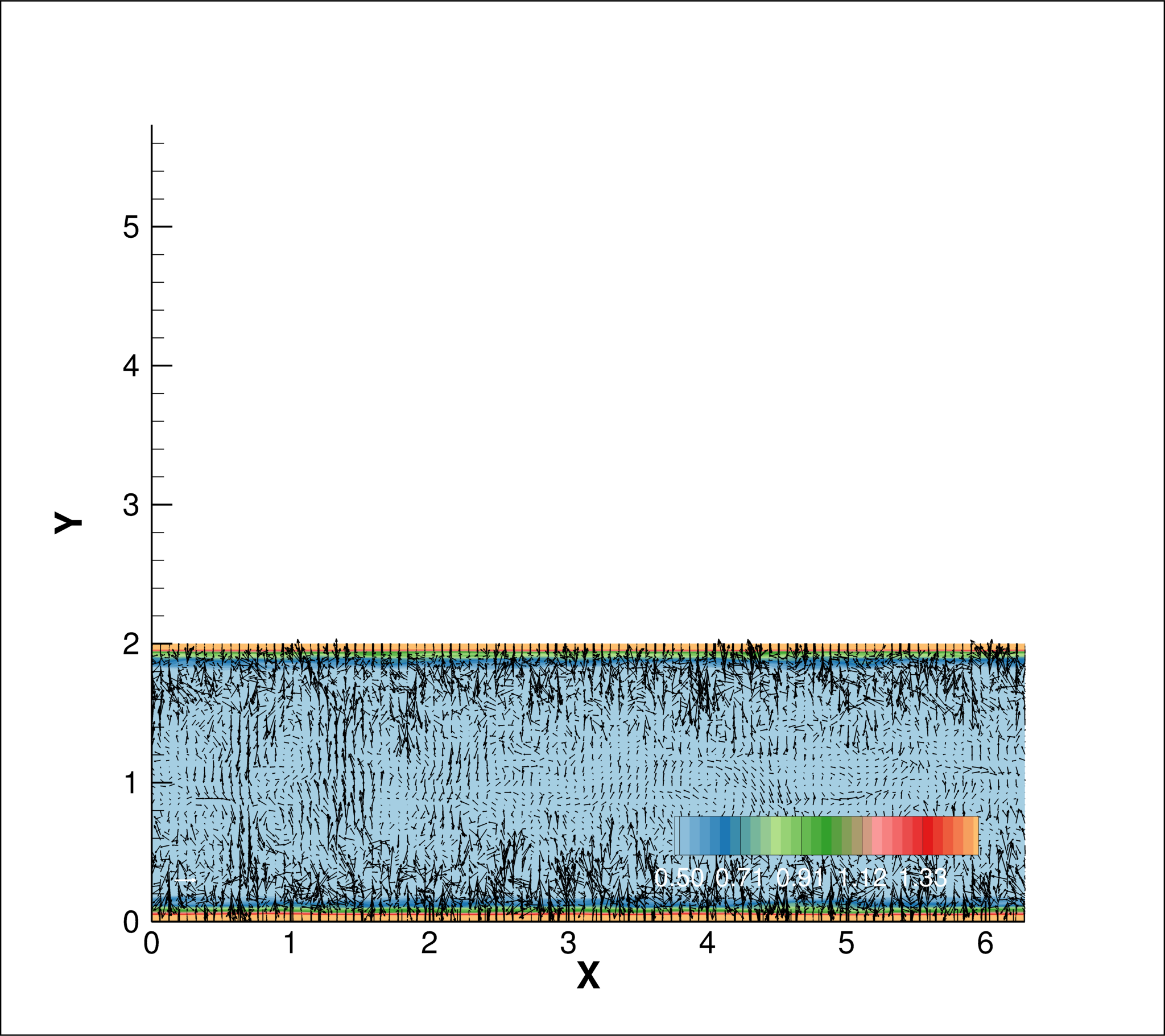}
\put(7.0,8.0){\color{white}\rule{4.2pt}{2pt}}
\put(-4.0,9.0){\color{black}\rotatebox{90}{0.4}}
\end{overpic}
\includegraphics[trim={7cm 5cm 10cm 48.5cm},clip=true,width=0.45\linewidth]{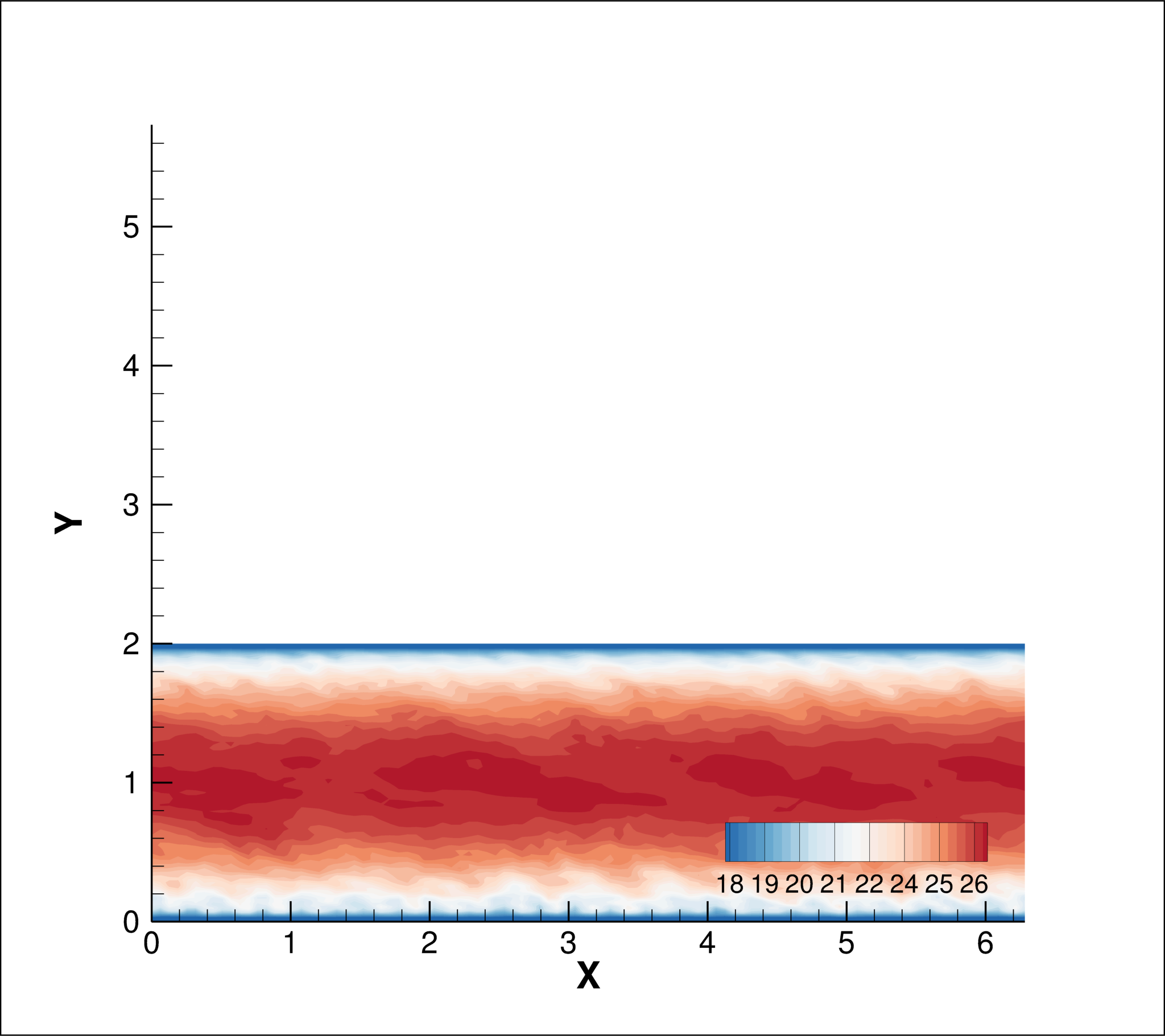}
\begin{overpic}
[trim={7cm 5cm 10cm 48.5cm},clip=true,width=0.45\linewidth]{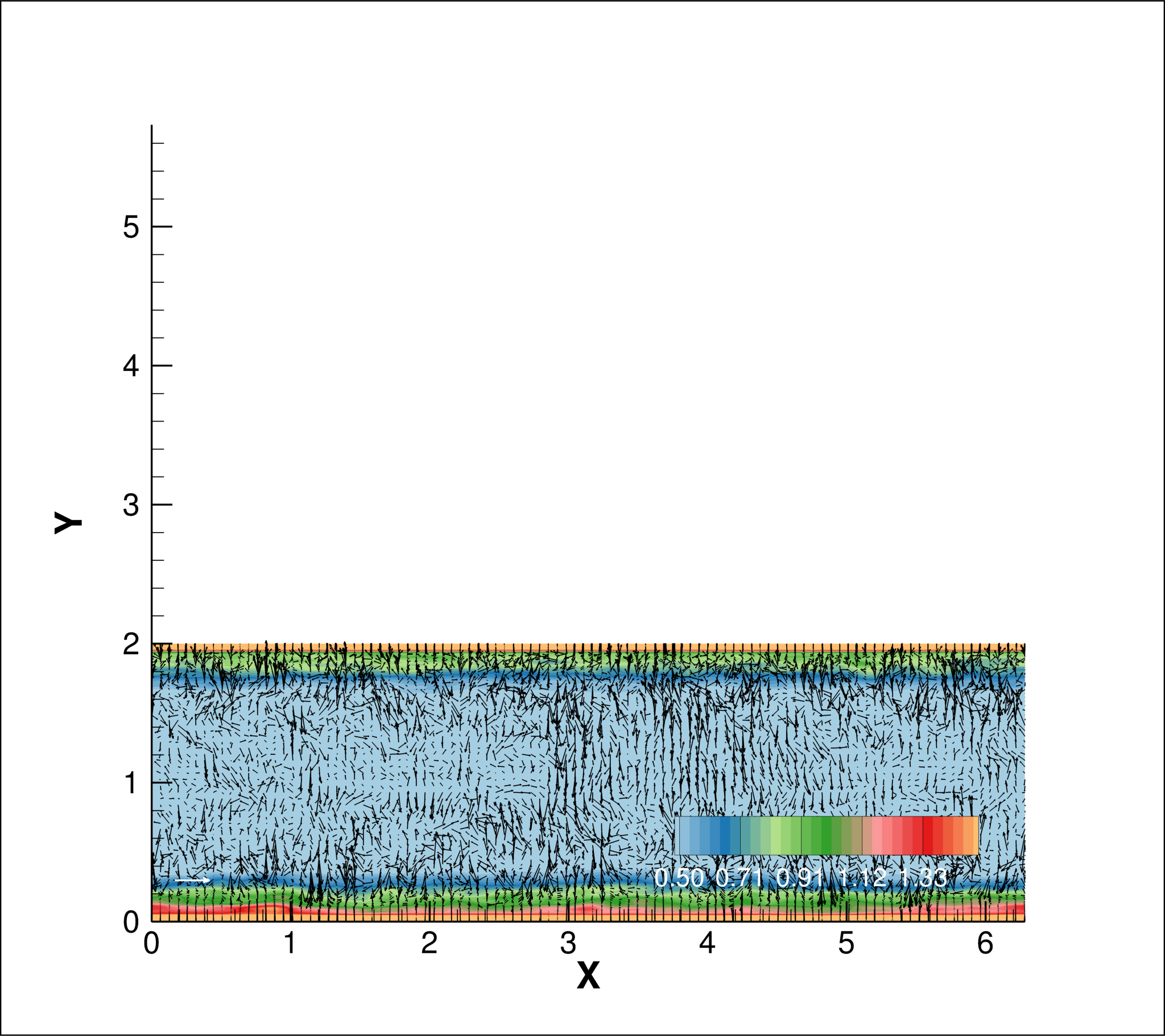}
\put(7.0,8.0){\color{white}\rule{7.7pt}{2pt}}
\put(-4.0,9.0){\color{black}\rotatebox{90}{0.8}}
\end{overpic}
\includegraphics[trim={7cm 5cm 10cm 48.5cm},clip=true,width=0.45\linewidth]{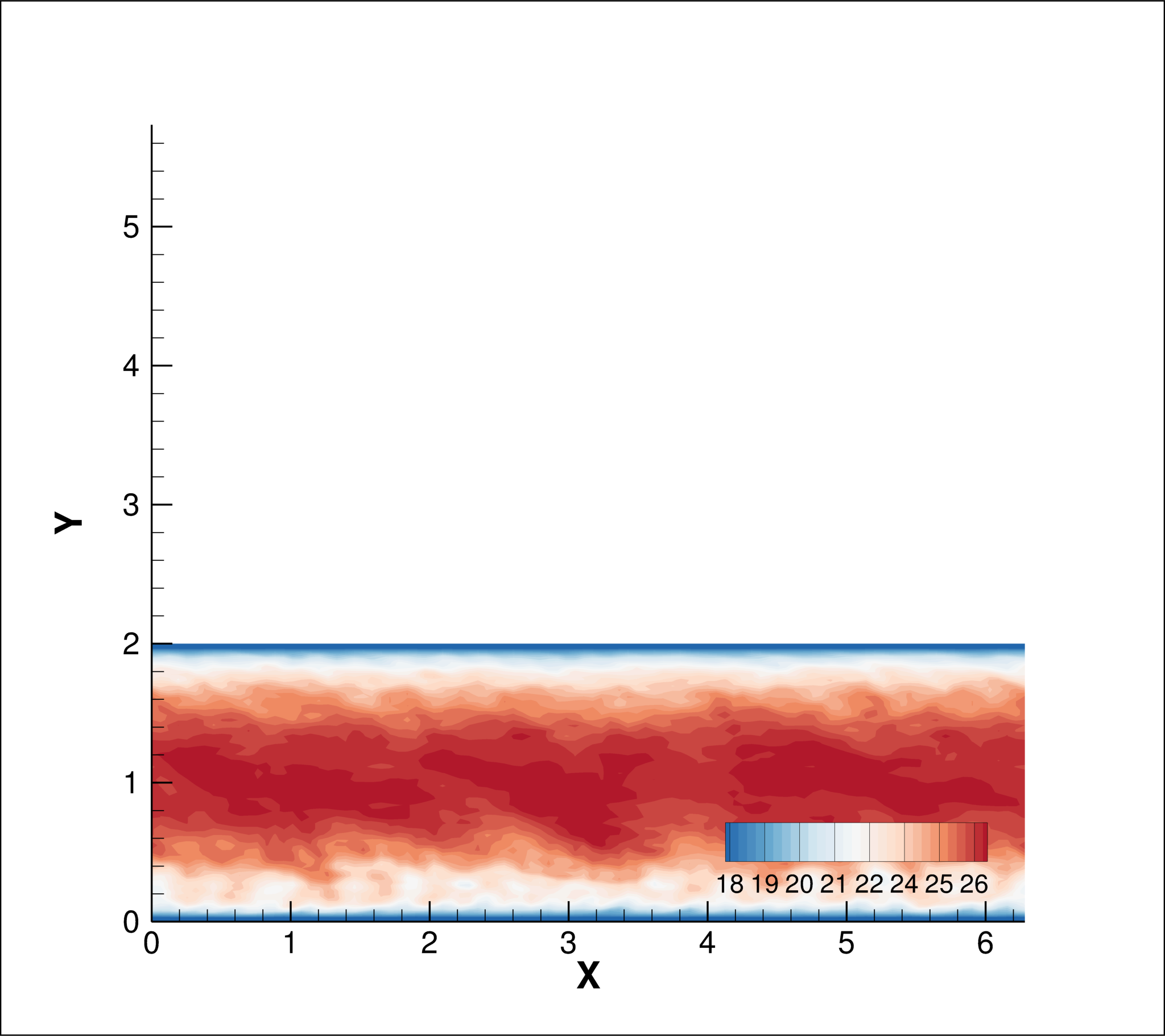}
\begin{overpic}
[trim={7cm 5cm 10cm 48.5cm},clip=true,width=0.45\linewidth]{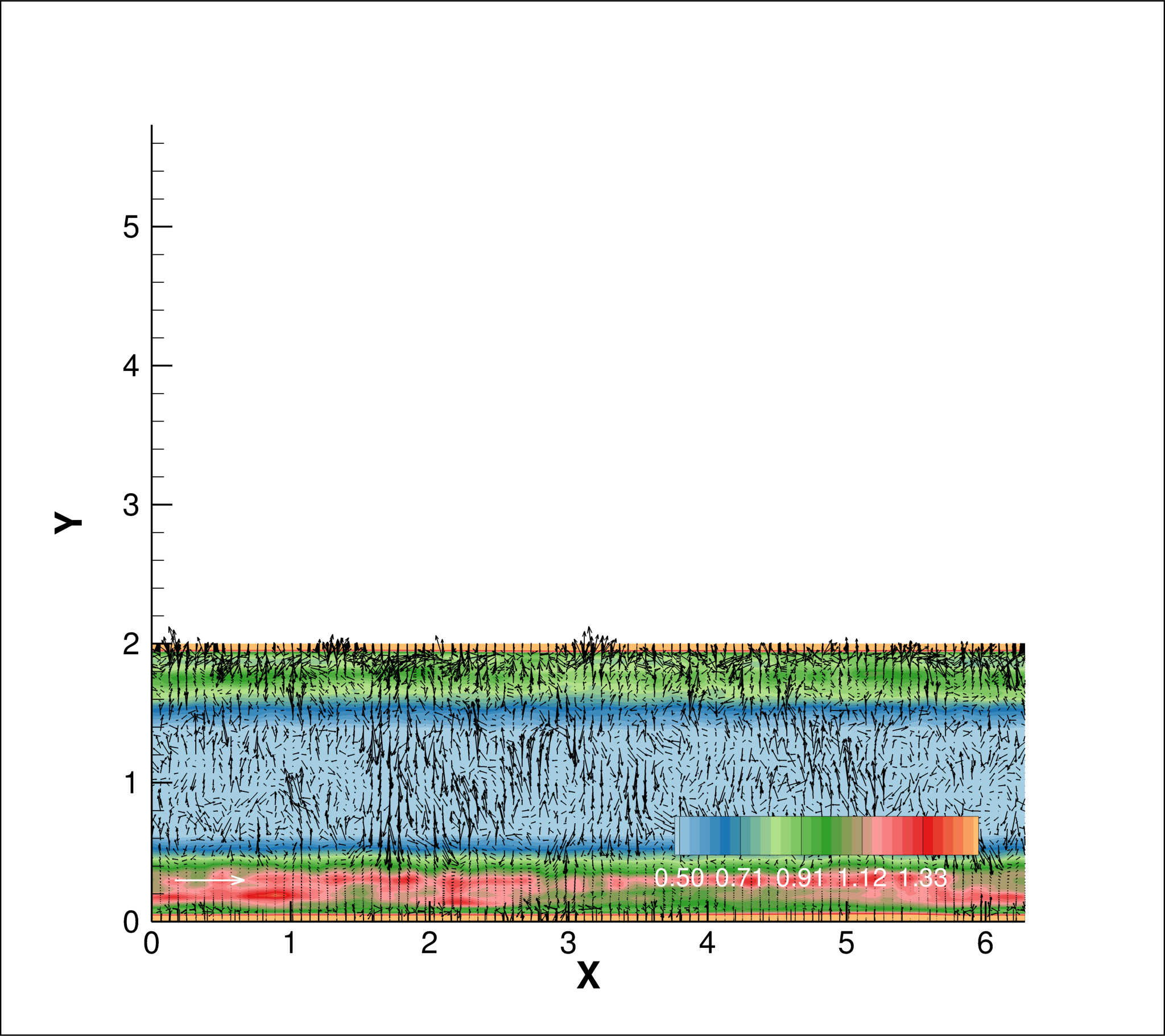}
\put(7.0,8.0){\color{white}\rule{15.7pt}{2pt}}
\put(-4.0,9.0){\color{black}\rotatebox{90}{1.6}}
\end{overpic}
\includegraphics[trim={7cm 5cm 10cm 48.5cm},clip=true,width=0.45\linewidth]{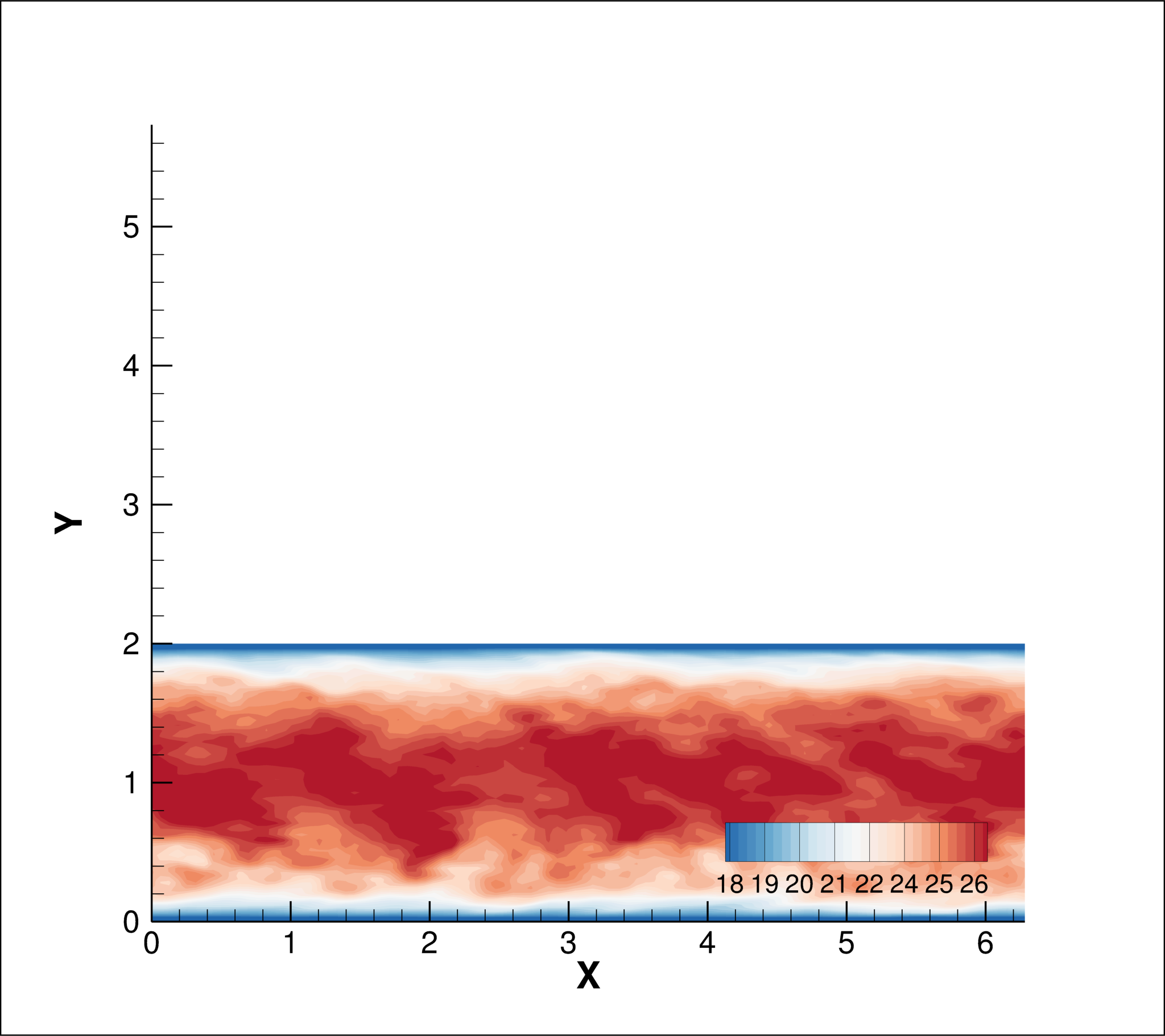}
\begin{overpic}
[trim={7cm 5cm 10cm 48.5cm},clip=true,width=0.45\linewidth]{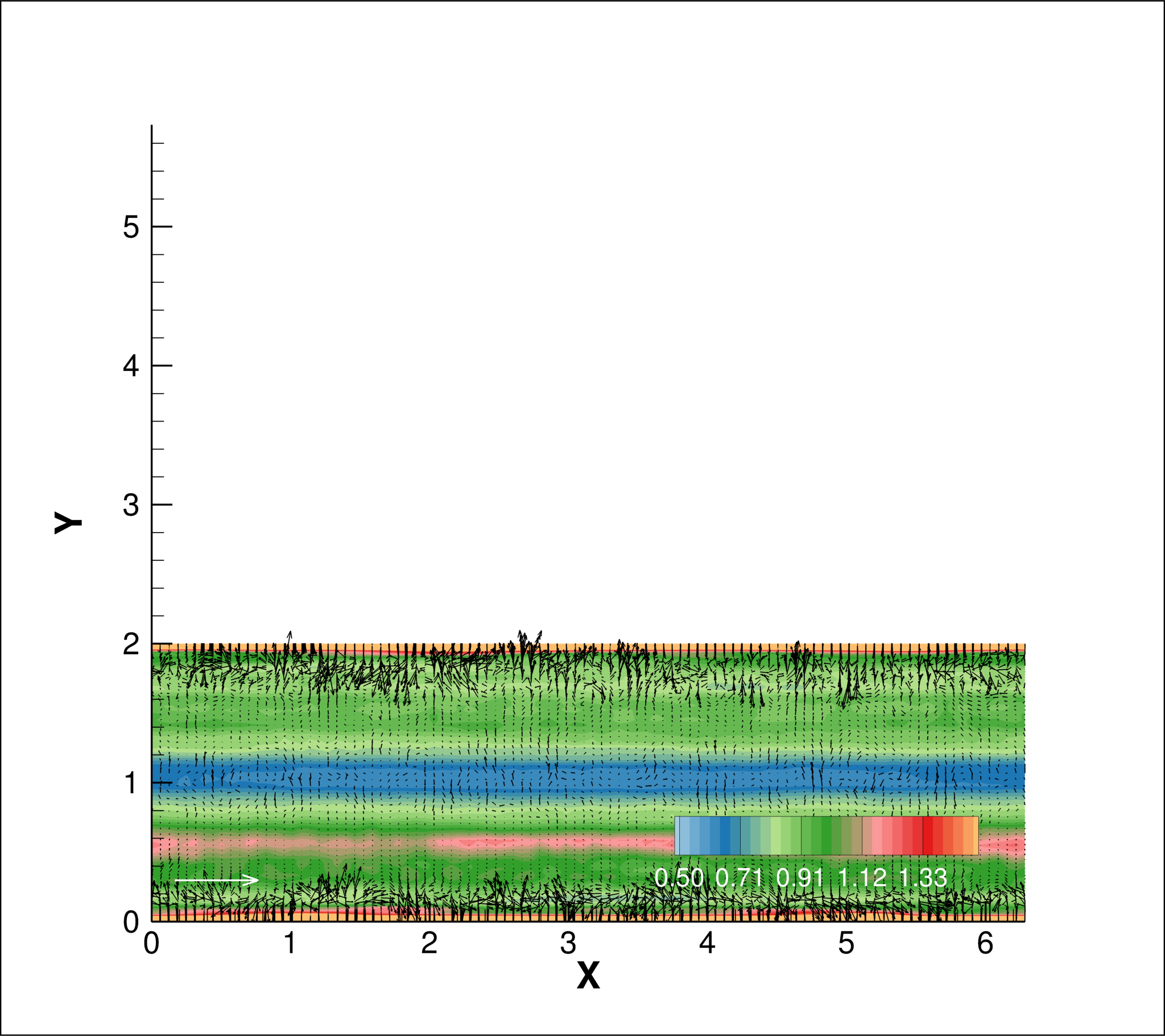}
\put(7.0,8.0){\color{white}\rule{19.0pt}{2pt}}
\put(-4.0,9.0){\color{black}\rotatebox{90}{3.2}}
\end{overpic}
\includegraphics[trim={7cm 5cm 10cm 48.5cm},clip=true,width=0.45\linewidth]{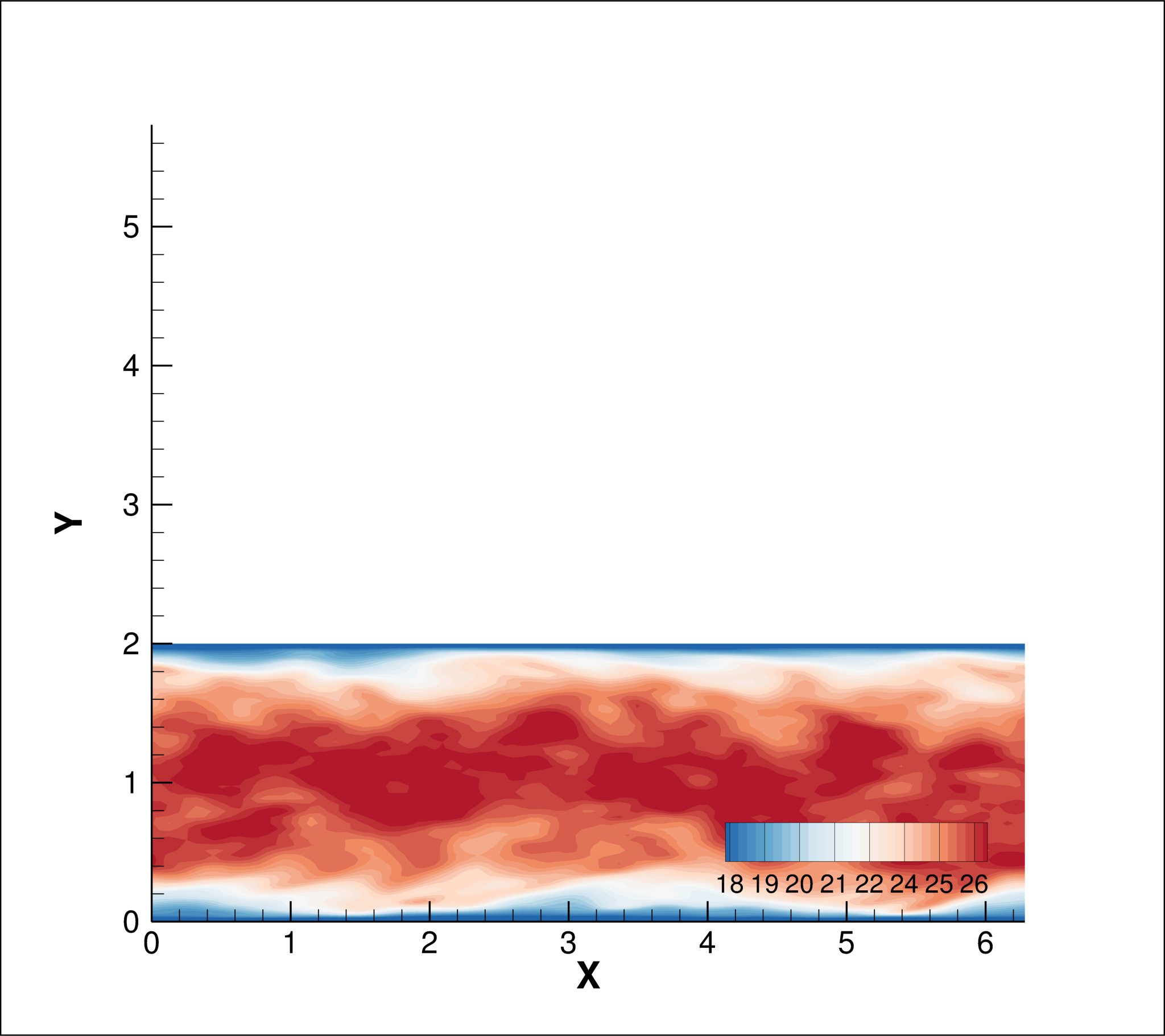}
\begin{overpic}
[trim={7cm 5cm 10cm 48.5cm},clip=true,width=0.45\linewidth]{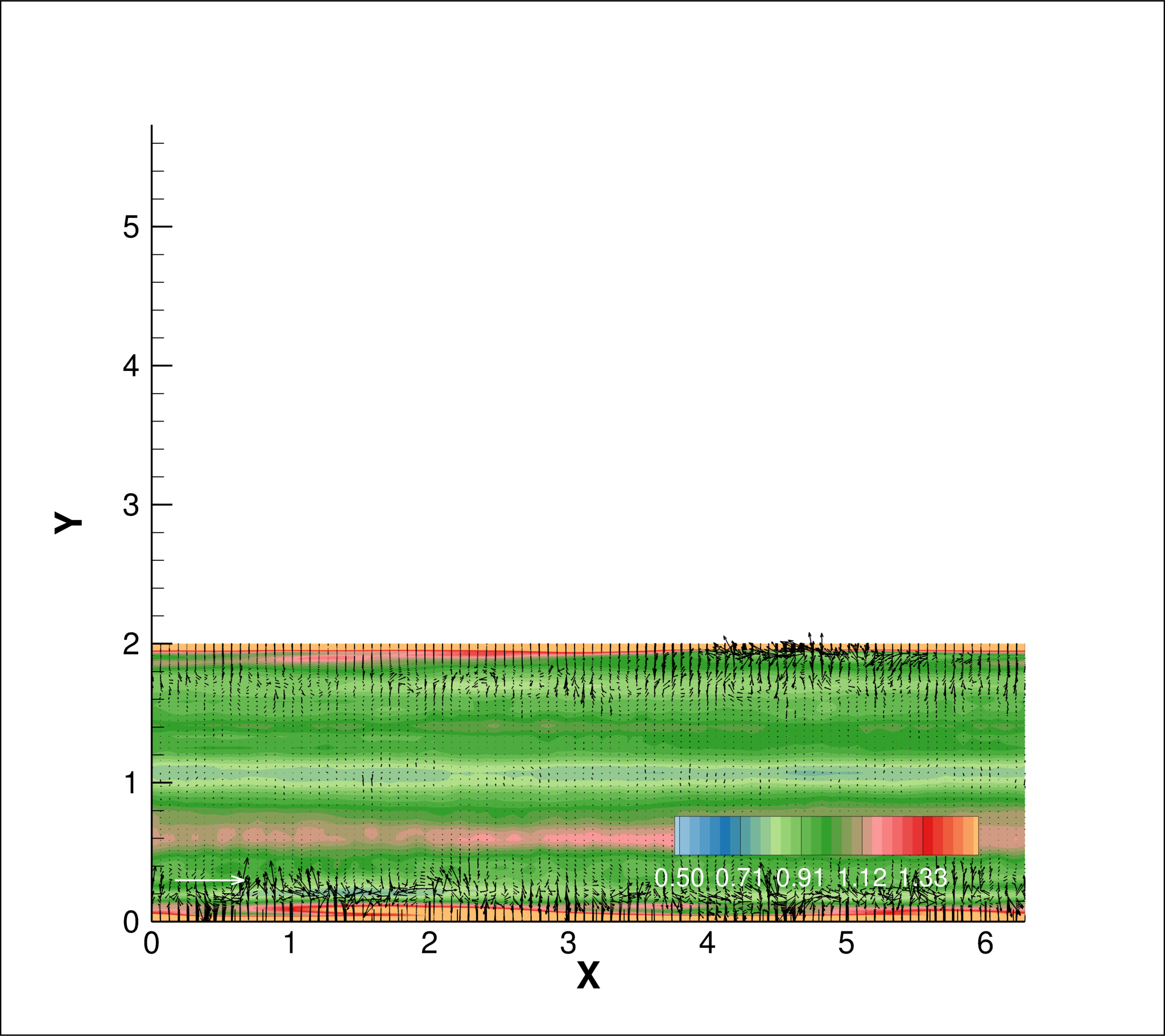}
\put(7.0,8.0){\color{white}\rule{19.0pt}{2pt}}
\put(-4.0,9.0){\color{black}\rotatebox{90}{4.0}}
\put(50.0,-5.0){\color{black}\rotatebox{0}{$\la{}r_\mathcal{M}\ra$}}
\end{overpic}
\begin{overpic}
[trim={7cm 5cm 10cm 48.5cm},clip=true,width=0.45\linewidth]{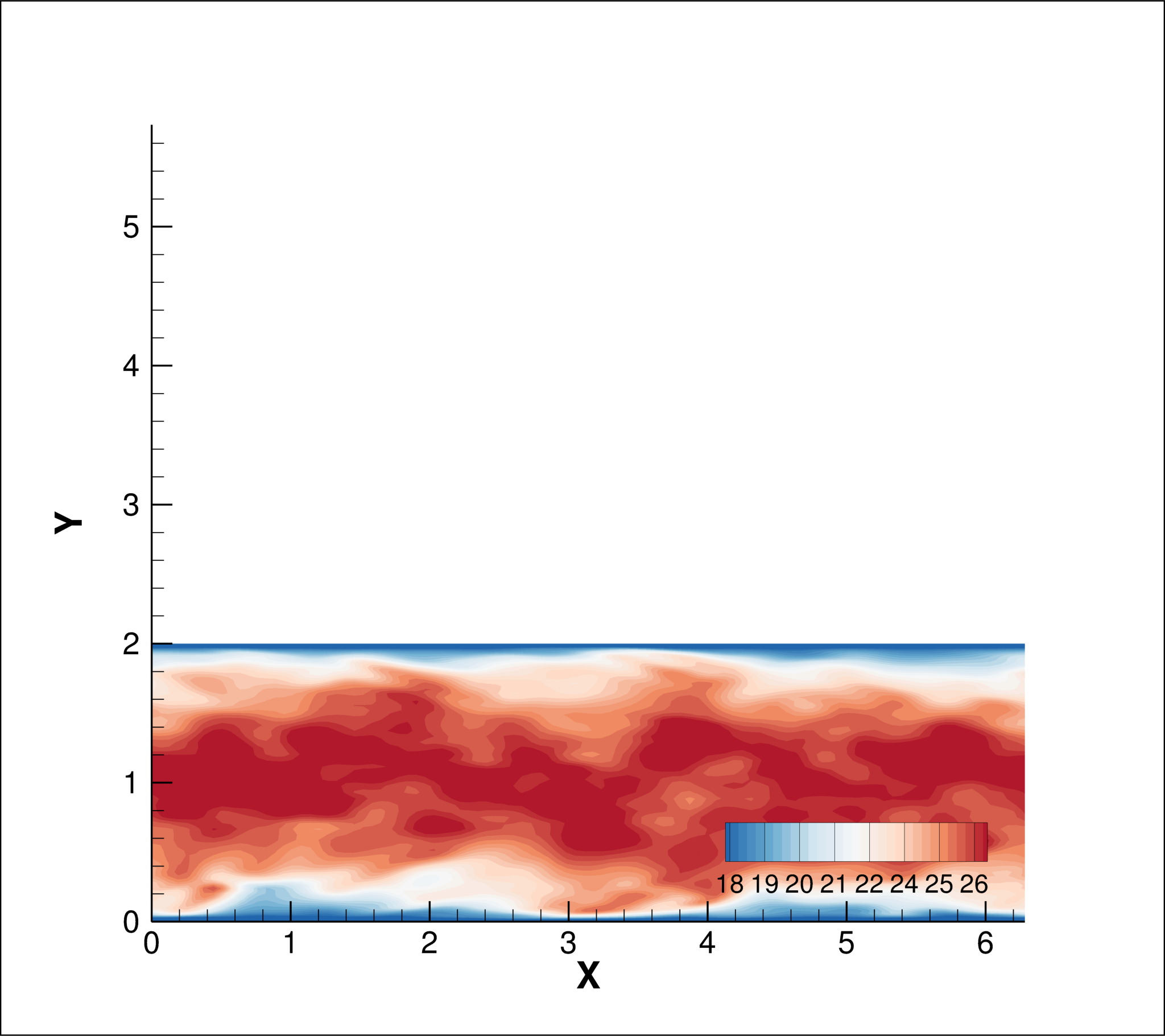}
\put(50.0,-5.0){\color{black}\rotatebox{0}{$\overbar{u_x}$}}
\end{overpic}
\end{center}
\caption{Snapshots of mean resolution adequacy parameter $\la{}r_\mathcal{M}\ra$ (\ref{rd}) and total 
streamwise velocity for the proposed hybrid method applied to fully developed channel flow at 
$Re_\tau\approx5200$ with forcing vector field, $F_i$, overlaid.  Time increments are labeled with 
mean velocity flow throughs.  Vector field scale indicated by each respective white marker bar of 
constant magnitude $(k_{max})^{1/2}/T_{max}$ where the ``max'' subscript indicates the maiximum value 
over the entire domain.  Half-channel shown here.}
\label{fig:rd_ux}
\end{figure}
\begin{figure}[htp!]
\begin{center}
\begin{subfigure}{0.45\linewidth}
\begin{center}
\includegraphics[width=0.85\linewidth]{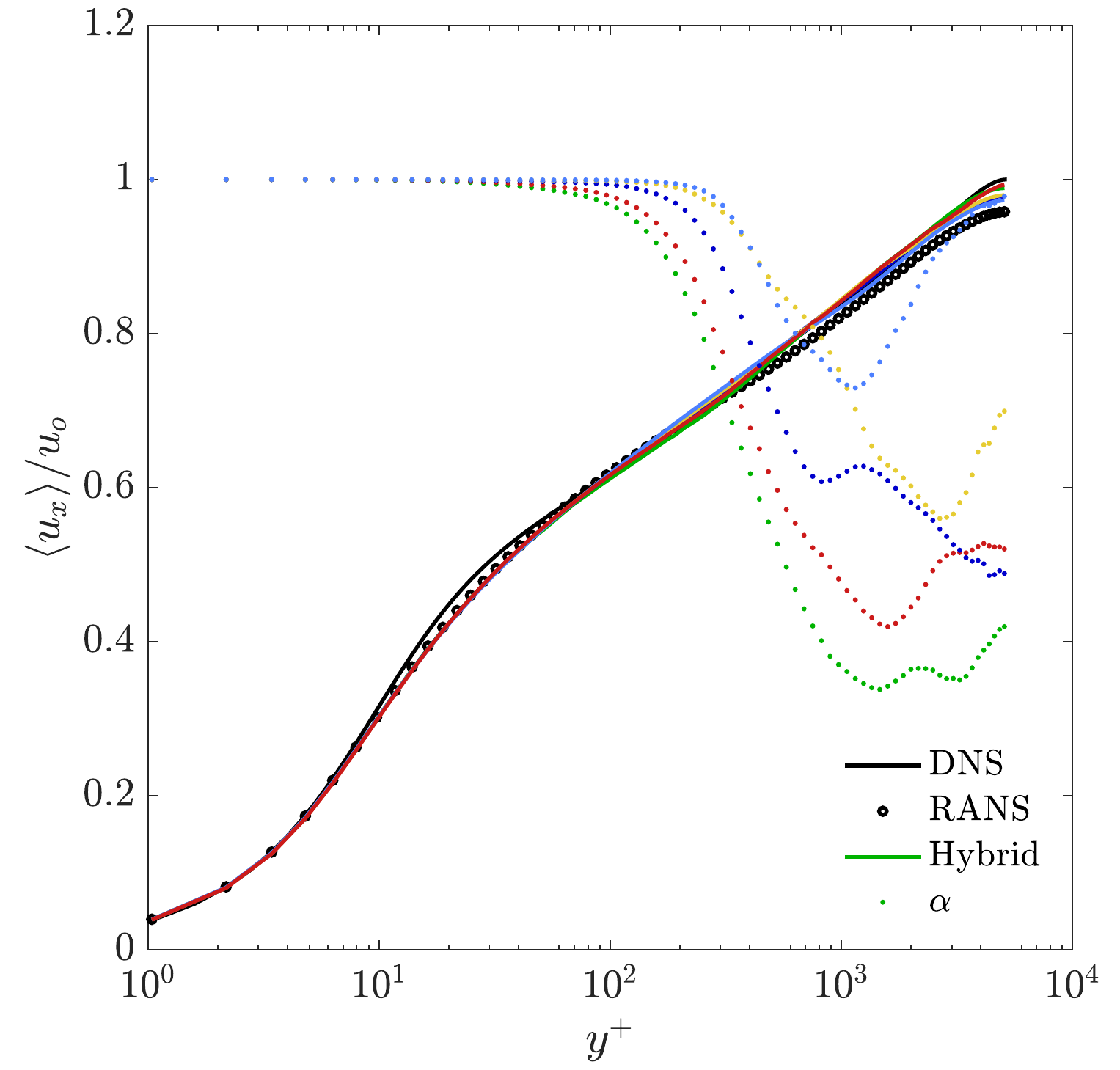}
\caption{Volume averaged $\bar{u}_x$ and $\alpha$}\label{mux}
\end{center}
\end{subfigure}
\begin{subfigure}{0.45\linewidth}
\begin{center}
\includegraphics[width=0.85\linewidth]{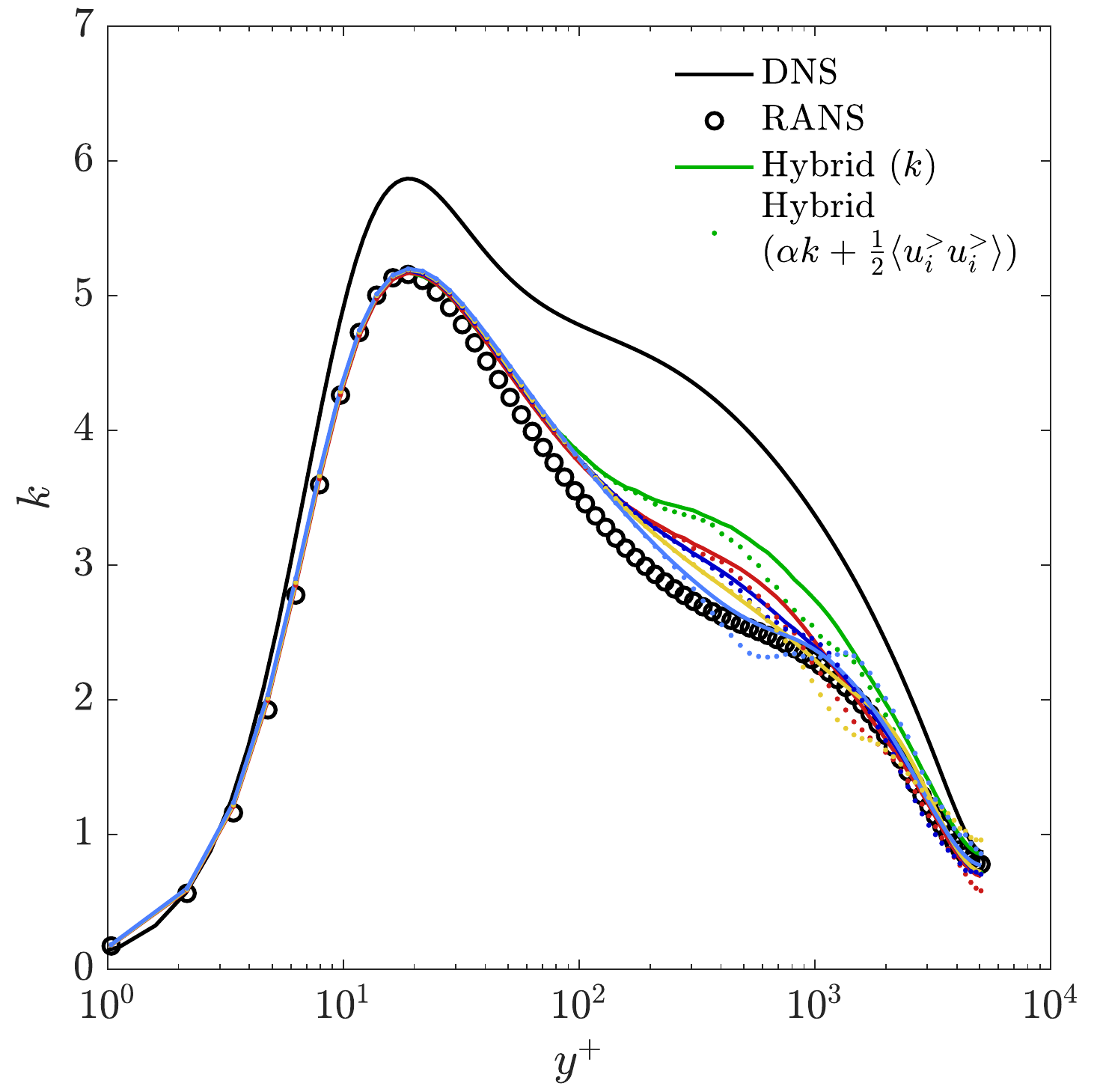}
\caption{Turbulent kinetic energy}\label{mtke}
\end{center}
\end{subfigure}
\end{center}
\caption{Normalized mean streamwise velocity for fully developed channel flow at $Re_\tau\approx5200$ with overlaid level of resolved turbulence (a) and turbulent kinetic energy at the steady-state time (b).  Volume averaged at times:  \textcolor{lblue}{---} $0.8$, \textcolor{yellow}{---} $2.4$, \textcolor{blue}{---} $4.0$, \textcolor{red}{---} $8.0$, and \textcolor{green2}{---} $40$ flow-throughs.}
\label{fig:poc}
\end{figure}

For the first case, the hybrid simulation is initialized from a
steady-state RANS solution.  At $t_f = 0$, the forcing is activated,
and the state evolves in time from the RANS initial condition to a
statistically stationary hybrid state.  This evolution is summarized
in Figures~\ref{fig:rd_ux} and~\ref{fig:poc}.

%
During the first step of forcing (Fig.\ref{fig:rd_ux}, $t_f=0.0$) the
resolved field is simply the RANS state and a large core of excess
resolution is present as indicated by $\la{}r_\mathcal{M}\ra<1$.
Without any resolved turbulence, $F_i$ is able to aggressively
generate fluctuations.  Note that the vector field scaling bar
lengthens as time progresses indicating equivalent plotted vector
lengths are decreasing in magnitude.  A short time later
(Fig.\ref{fig:rd_ux}, $t_f=0.4$), fluctuations of varying scales are
present, and the forcing is reduced in response to the existing
fluctuations limiting where energy can be added provided the
prescribed TG structure.  This trend continues through each successive
snapshot while a front of $\la{}r_\mathcal{M}\ra\approx1$ expands away
from the wall indicating a growing region of grid-level LES.  Finally,
after multiple flow-throughs (Fig.\ref{fig:rd_ux}, $t_f=4.0$), a wide
spectrum of resolved fluctuations is clearly present with grid-level
LES having reached the center of the channel.  Throughout the
simulation, $\la{}r_\mathcal{M}\ra$ remains large very near the wall
thereby maintaining locally pure RANS mode, as expected.

Large scale structures develop more slowly (compare $\overbar{u_x}$ at $t_f=0.4$ and $t_f=3.2$, 
for instance).  The general idea of the proposed active forcing approach is to force the largest 
turbulent structures initially and gradually reduce the size of the forcing structure towards the 
grid length scale as those scales are populated with resolved turbulence.  The apparent deviation 
from the anticipated behavior indicates the scale of the TG should be increased initially, perhaps 
significantly.


Throughout the process of turbulence generation, the mean streamwise velocity profile 
remains accurate (Fig. \ref{mux}). The expected logarithmic profile is observed at all 
stages of development with no major deviations from the DNS data.  This result is quite  
remarkable as for varying levels of resolved turbulence, most of which are not a 
grid-resolved LES, there is no apparent MSD or log-layer mismatch.  Hence, the 
gradual body forcing approach, despite the crude form used here, appears to be 
sufficient for generating fluctuations that yield realistic behavior.  

Results for the TKE are also encouraging (Fig.~\ref{mtke}).  The RANS
$\overbar{v^2}$-$f$ result for $k$ significantly under-predicts the DNS
result over most of the channel.  Hybrid results are generally seen to 
move towards DNS in the region where some turbulence is resolved.  
The total turbulent kinetic energy from the RANS transport equation, $k_{tot}$,  
improves in response to a more accurate production term due to the 
contribution from the mean of the resolved stress.  Comparing $k_{tot}$ to the TKE computed from the model and the resolved field---i.e., $k_{sgs} + k_{res} = \alpha k_{tot} + \la u^>_i u^>_i \ra/2$---shows slight
disagreement early in the forcing.  However, as 
the simulation progresses, this modeled-resolved inconsistency decreases 
to only a small amount around $y^+\approx1000$.  This region corresponds 
to the slightly under-resolved band ($\la{}r_\mathcal{M}\ra>1$) in 
Fig. \ref{fig:rd_ux} at $t_f=4.0$ and will be examined further in later work.

\begin{figure}[htp!]
\begin{center}
\begin{subfigure}{0.45\linewidth}
\begin{center}
\includegraphics[width=0.85\linewidth]{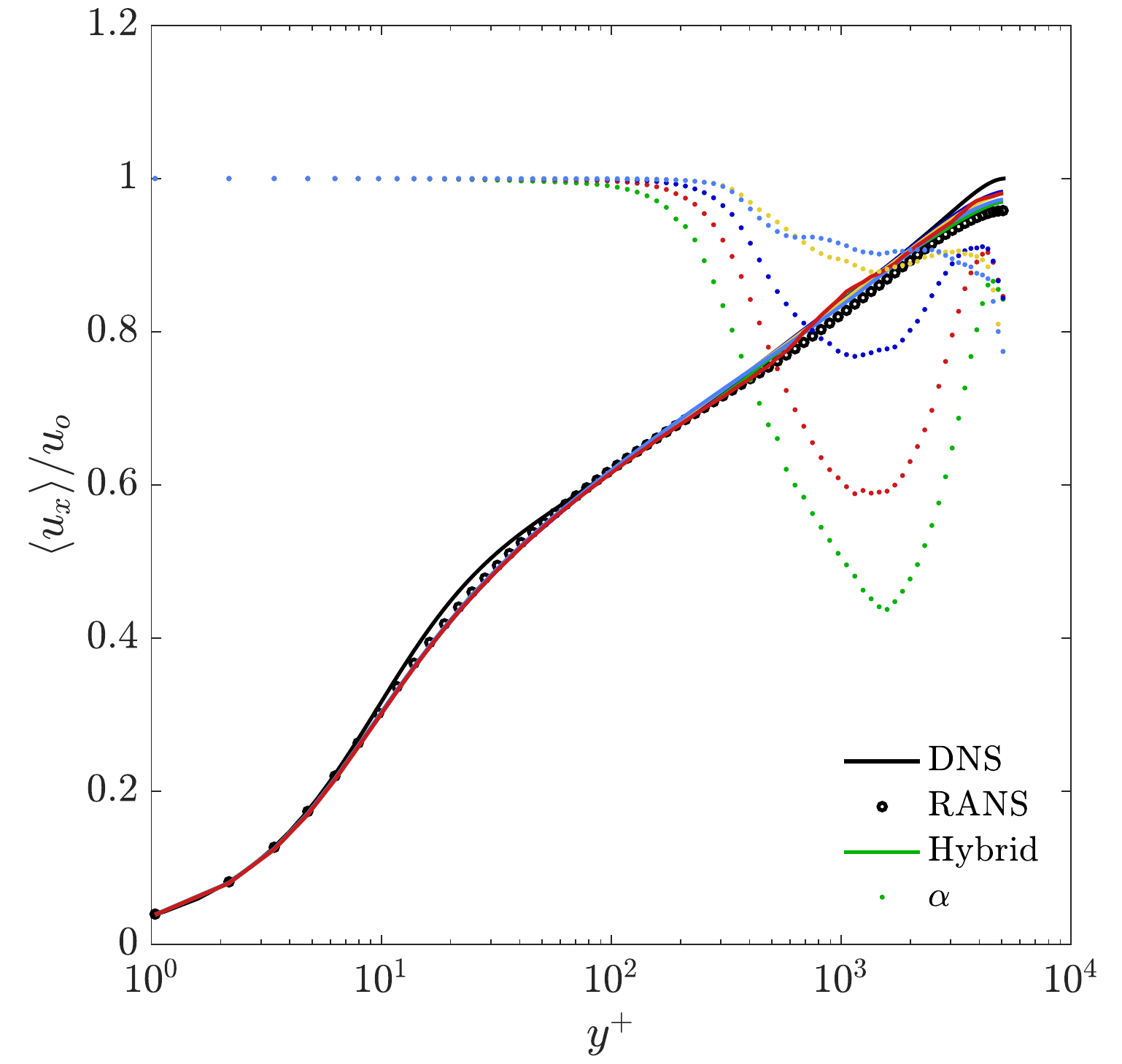}
\caption{Plane averaged $\bar{u}_x$ and $\alpha$}\label{s_mux}
\end{center}
\end{subfigure}
\begin{subfigure}{0.45\linewidth}
\begin{center}
\includegraphics[width=0.85\linewidth]{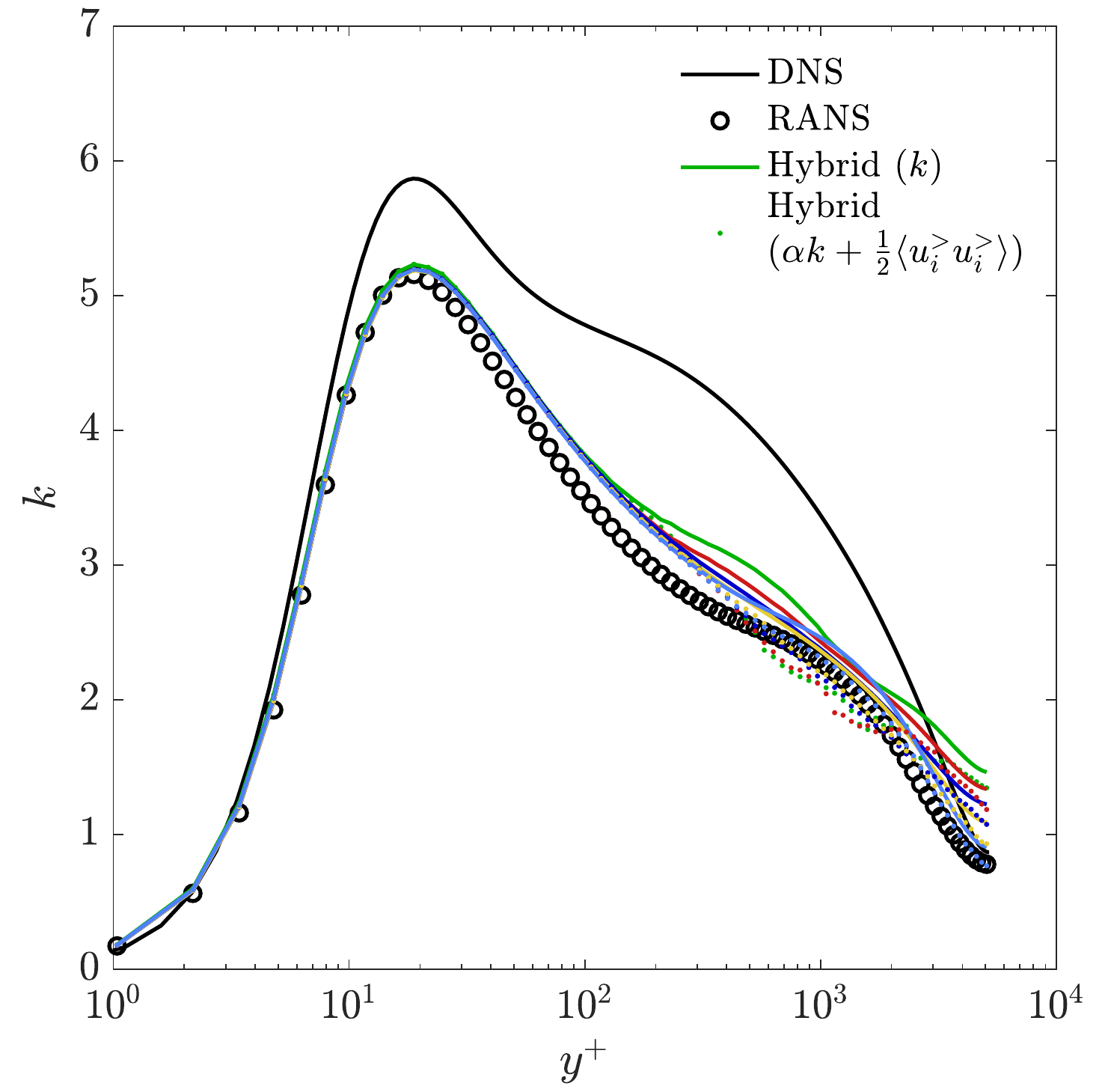}
\caption{Turbulent kinetic energy}\label{s_mtke}
\end{center}
\end{subfigure}
\end{center}
\begin{center}
\begin{overpic}
[trim={0.2cm 29.2cm 0.2cm 29.2cm},clip=true,width=0.95\linewidth]{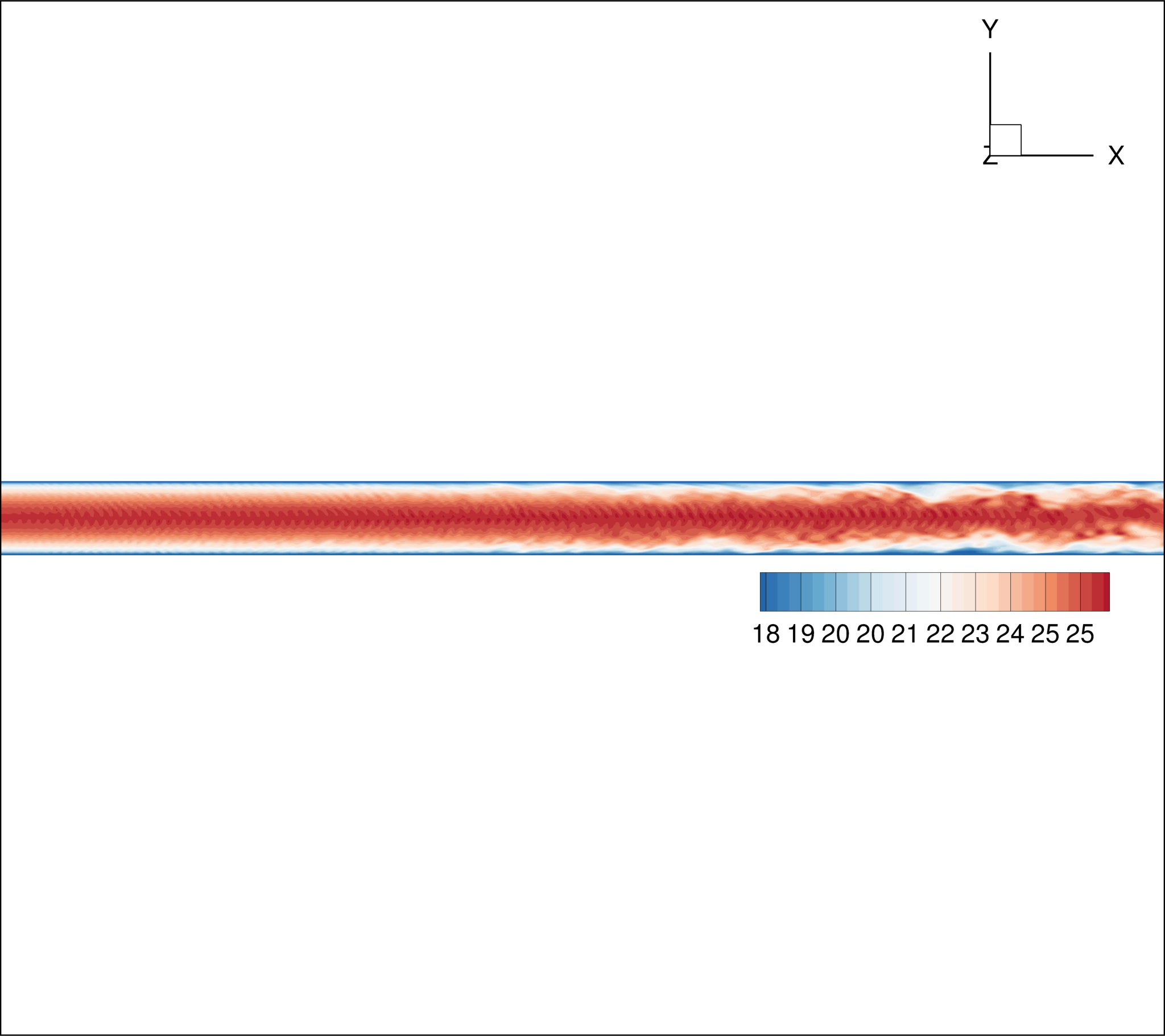}
\put(10.0,1.0){\color{lblue}\rule{1pt}{27.0pt}}
\put(30.0,1.0){\color{yellow}\rule{1pt}{27.0pt}}
\put(50.0,1.0){\color{blue}\rule{1pt}{27.0pt}}
\put(70.0,1.0){\color{red}\rule{1pt}{27.0pt}}
\put(90.0,1.0){\color{green2}\rule{1pt}{27.0pt}}
\put(-2.6,2.6){\color{black}\rotatebox{90}{$\overbar{u_x}$}}
\end{overpic}
\begin{overpic}
[trim={0.2cm 29.2cm 0.2cm 29.2cm},clip=true,width=0.95\linewidth]{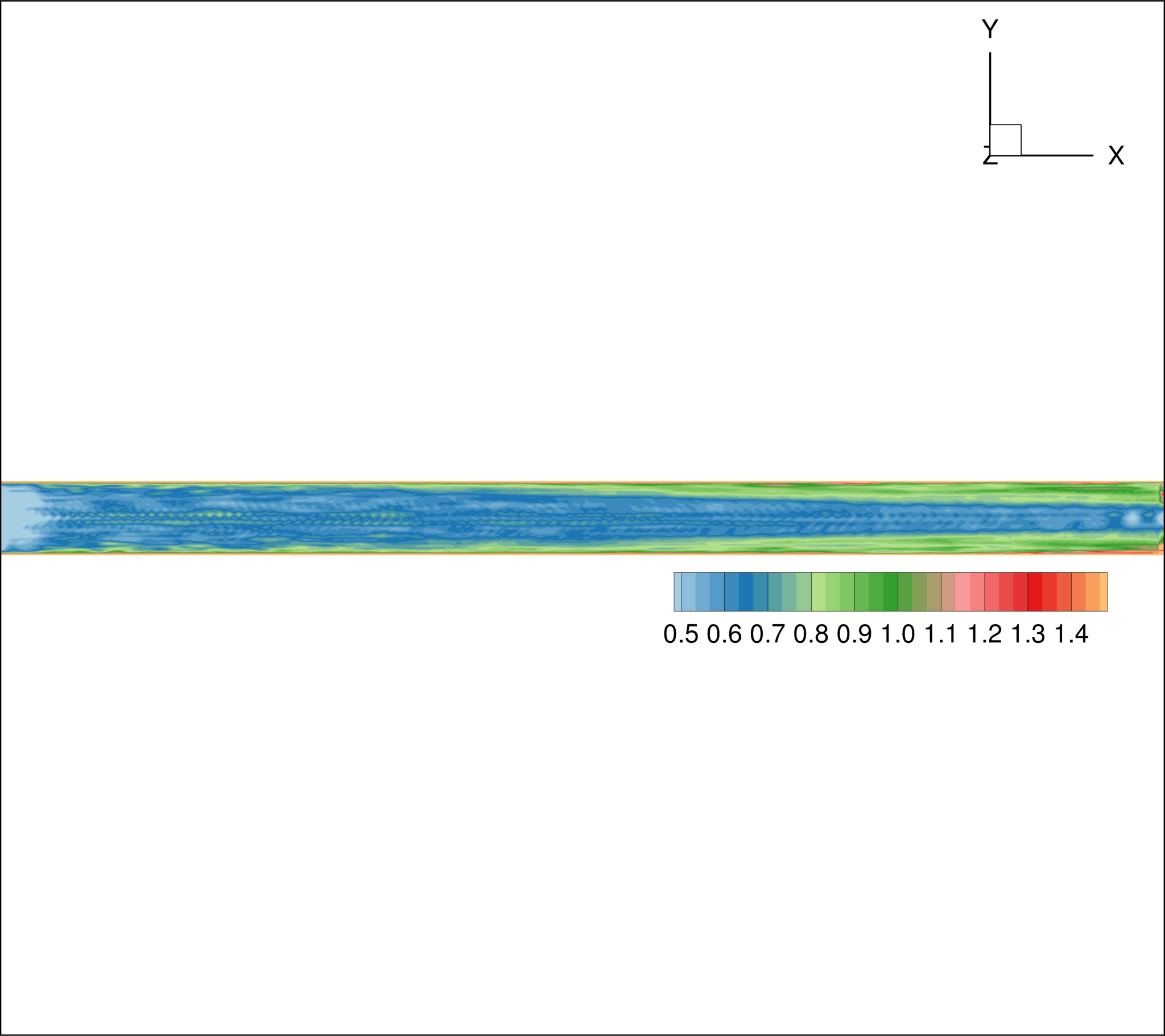}
\put(-3.0,1.5){\color{black}\rotatebox{90}{$\la{}r_\mathcal{M}\ra$}}
\end{overpic}
\end{center}
\caption{Normalized mean streamwise velocity for channel flow at $Re_\tau\approx5200$ with overlaid level of resolved turbulence (a) and TKE (b).  Plane-averaged at positions:  \textcolor{lblue}{---}$\pi$, \textcolor{yellow}{---} $3\pi$, \textcolor{blue}{---} $5\pi$, \textcolor{red}{---} $7\pi$, and \textcolor{green2}{---} $9\pi$ as indicated in the contour plot of $\overbar{u_x}$.  Both $\overbar{u_x}$ and $\la{}r_\mathcal{M}\ra$ use the same colormap as in Fig. \ref{fig:rd_ux}}
\label{fig:spatial}
\end{figure}
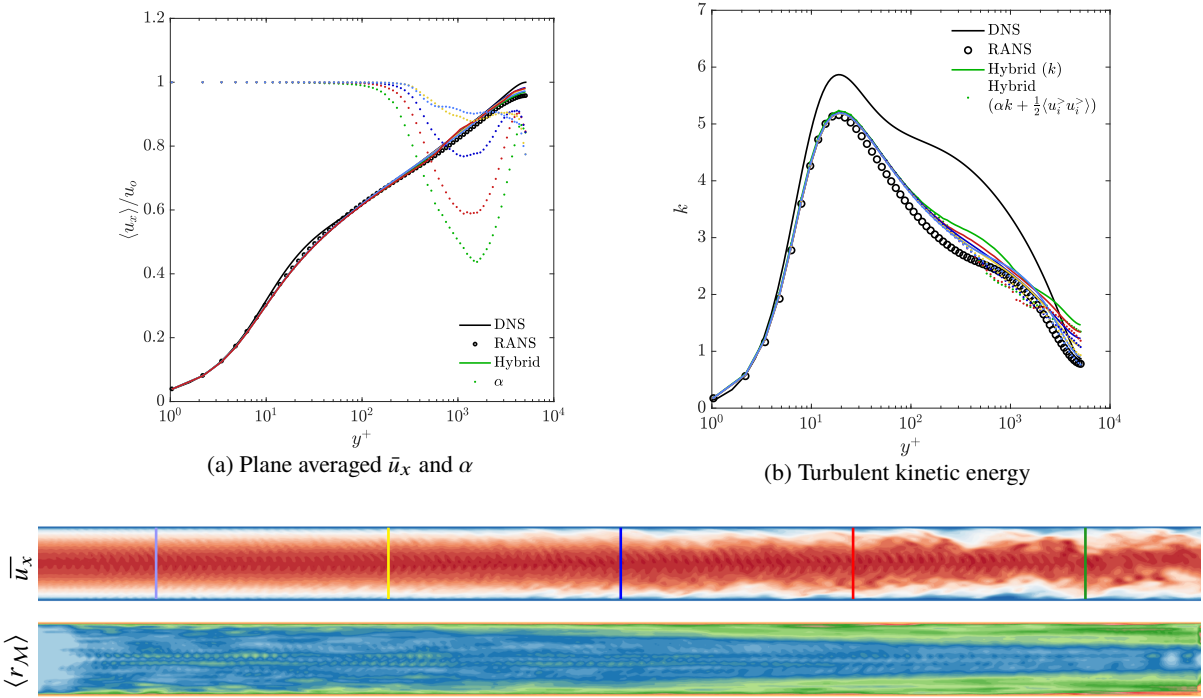

Next, we examine the spatially developing characteristics of the
hybrid method by extending the length of the channel to $10\pi$ and
using a RANS inlet condition with convective outflow.  Thus, this test
case is more representative of typical applications.  The grid spacings
in plus units has been coarsened to $\Delta^+\approx450$ in the span 
and streamwise directions.  The results are summarized in Figures~\ref{fig:spatial}.

Again, for all spatial locations and levels of resolved turbulence, the expected 
logarithmic profile in the mean streamwise velocity is attained (Fig. \ref{s_mux}).  
A very minor deviation is observed at the $7\pi$ location.  Resolved turbulence 
develops more gradually in the core of the channel in comparison to around 
$y^+\approx1000$.  This is likely due to the magnitude of the forcing scaling with 
the local TKE and time scale (Eq. \ref{Ftar}).  The under-utilized resolution is 
apparent in accompanying $\la{}r_\mathcal{M}\ra$ contour plot where the grid-resolved LES 
``front'' has not extended to the channel center by the outlet. Naturally, it is highly desirable for 
the model to transition from RANS to grid-resolved LES with as little time and 
space as possible.  Therefore, improving slow-core development is another avenue 
for improving future forcing formulations.  Contrary to results in the temporally developing case (Fig. 
\ref{mtke}), the TKE from both the RANS model and accounting for the contribution 
from the resolved, field exceeds the DNS value in the center of the channel (Fig. 
\ref{s_mtke}).  The reason for this inconsistency is not apparent though it does not 
seem to corrupt the mean flow.  

These preliminary results indicate that the model-split hybridization with active
forcing and SGET anisotropy is promising and worthy of further study
and development.

\section{Conclusions}
\label{sec:conclusions}

By examining the failures of recently developed models, three general deficiencies 
of existing hybrid methods have been identified: use of a single eddy viscosity to 
represent both subgrid stress and energy transfer, the use of RANS-based transport 
models with fluctuating quantities, and the reliance on passive self-generation of 
resolved turbulence in hybrid regions.  To remedy these, a new modeling framework 
is proposed.  The novel aspects of the framework include 1) the model-split 
hybridization approach, 2) a forcing formulation which actively injects fluctuations in 
regions that are capable of supporting more turbulence, and 3) the ability to naturally 
incorporate anisotropic models, like the M43 SGET model.  Preliminary results for 
channel flow indicate that the model leads to good predictions of the mean velocity
profile for a wide range of resolutions and levels of resolved turbulence including 
regions of transitional, non-equilibrium developing LES.  While issues related to the 
slow production of resolved turbulence in the channel core and slight DNS-hybrid TKE 
core mismatch in the spatially developing case have been identified, the approach 
does seem to be a solution for the common hybrid issues of log-layer mismatch and MSD.

With initial results being very encouraging, significant evaluations
of more complex flow scenarios are warranted.  These include common
aerodynamic test cases with smooth wall separation and reattachment.
Further, unsteady flows where the current time averaging procedure may
prove problematic are of particular interest.  In addition, many 
avenues for potential refinement and improvement of the TAMS approach are apparent.  
For example, the prescribed forcing structure may be enhanced to more closely mimic real turbulence 
to allow more rapid transfer of energy from modeled to resolved scales;  stress anisotropy 
may be included to improve model performance in regions of only small amounts of 
resolved turbulence; the effects of turbulence anisotropy, instead of only resolution 
anisotropy, may be added with more advanced SGET models;  and the combination 
of various RANS and SGET models may be exploited to leverage their individual strengths.  
Though early in its development, the proposed TAMS framework is a very promising 
approach which seems capable of achieving the hybrid RANS/LES goals of a robust, 
predictive, and cost-effective method of simulating complex fluid flows.

\section*{Appendix}
\label{sec:appendix}

\noindent\textbf{{I. Previous Models}}\\
Details of previous models whose failures were used to guide the requirements 
for the current TAMS approach are briefly detailed here. Models used to
show the two opposite failure modes in the WMH (Fig. \ref{fig:hump_flow}) 
retain the ``Model A'' and ``Model B'' distinctions used in the figure while the model used 
to illustrate the effects of including fluctuating terms in RANS equations (Fig. 
\ref{fig:chan_expected_sources}), is labeled ``Model C''.  All three 
relied on passive self-generation of resolved turbulence.\\*

\noindent\textbf{\emph{Model A:}}
This model was the earliest version which used an anisotropic resolution indicator 
to guide hybridization and was loosely based on the work of Perot and Gadebusch 
\cite{perot:2007}.  Overall hybridization was achieved with the hybrid parameter, 
$\alpha$, (not the same definition as used in TAMS)
  \begin{equation}
  D_t\overbar{u}_i=-\tfrac{1}{\rho}\pd_i\overbar{p}+\pd_j({\alpha}\tau_{ij})+\nu\pd_k\pd_k\overbar{u}_i
  \end{equation}
  \begin{equation}
  D_tk_m={\alpha}\mathcal{P}_k-\varepsilon+\pd_i\Big{[}\Big{(}\nu+\tfrac{\nu_t}{\sigma_k}\Big{)}\pd_ik_m\Big{]}.
  \end{equation}
This parameter was advanced locally as 
  \begin{equation*}
  \frac{d\alpha}{dt}=\frac{1}{T_m}(r_k-1)
  \end{equation*}
in response the initial energy-based resolution adequacy parameter
\begin{equation}
r_k 
= 
\frac{\frac{2}{3} C_F \lambda^{Q}_{max}}{\overline{v^2}},
\label{rk}
\end{equation}
%
where $C_F$ is a constant which related the structure function to the subgrid kinetic 
energy with $\tilde{Q}_{ij}$ being a measure of the anisotropic second-order structure function
\begin{equation}
\tilde{Q}_{ij}=\mathcal{M}_{im}Q_{mn}\mathcal{M}_{nj},
\label{Qtil}
\end{equation}
\begin{equation}
Q_{ij} = \partial_m{\bar{u}_k}\mcal{M}_{mi} \partial_n{\bar{u}_k} \mcal{M}_{nj},
\label{Qij}
\end{equation}
and $C_K$ is the Kolmogorov constant.  This resolution parameter attempted to compare 
the modeled $k_{sgs}$ with its estimate provided the information in the resolved field and 
resolution.  The hybrid stress made use of (\ref{tau_bar}) with an anisotropic eddy viscosity as 
\begin{equation}
\nu_{ij} = \nu_t a_{ij},
\quad
a_{ij} = (1 - \beta) \delta_{ij} + \beta \left( \frac{3}{\tilde{Q}_{kk}} \right) \tilde{Q}^{1/2}_{ij},
\quad
\beta = \min \left( \frac{1}{r_k}, 1 \right).
\label{eqn:anisotropic_eddy_visc_hybrid}
\end{equation}
In addition to the issue identified in \S\ref{sec:blending_problems} with using fluctuating terms in RANS-transport 
models, the use of a dissipation anisotropy for the entire stress, and ambiguous RANS/LES 
distinction, the filter-scale structure function measure in $Q_{ij}$ was found to only be 
valid in the limit of isotropic resolutions and gross under-estimated $k_{sgs}$.  Further, the 
construction of $Q_{ij}$ with the entire gradient is not valid in near-RANS regions.\\*

\noindent\textbf{\emph{Model B:}}
This model was a response to the shortcomings of Model A and treated the hybridization 
parameter as more of a dynamic model coefficient which modified the eddy viscosity in 
according to the resolution adequacy parameter.  Corrections to the anisotropic structure 
function for anisotropic grids were constructed with inertial range theory to arrive at an 
estimate of the expected unresolved fluctuations
\begin{equation}
\la{}u^<u^<\ra \approx \max\big{(}C_{sf}\tfrac{2}{3}\zeta_{ik}\zeta_{jl}\mathcal{Q}_{kl}\big{)},
\end{equation}
with $C_{sf}=0.367$, to be used for an anisotropic grid corrected resolution adequacy 
parameter where the anisotropic resolution correction was found using the $2/3$-law, 
\begin{equation}
\lambda^{\zeta}_{(i)} = C_{\zeta}\big{(}\lambda^{\hat{\mathcal{M}}}_{(i)}\big{)}^{1/3}\big{(}\lambda^{\hat{\mathbb{Q}}}_{(i)}\big{)}^{-1/2},
\label{zeta}
\end{equation}
where $\mathbb{Q}=\mathcal{Q}\varepsilon^{-2/3}$ and a ($\hat{\cdot}$) indicates normalization by 
the appropriate power of $\delta_{norm}=\min({\mathcal{M}})$.  Using Cayley-Hamilton theory, the 
nondimensional expected gradient-gradient product is expressed as
\begin{equation}
\ln\big{(}\lambda^{\hat{\mathbb{Q}}}_{i}\big{)}=\sum_{p=0}^2 g_{p}\ln(\lambda^{\hat{\mathcal{M}}}_{i})^{p}
\label{Q2}
\end{equation}
where the coefficients $g_i$, are function of the eigenvalues of $\mathcal{M}$.  
Similar to the M43 model, fourth-order polynomials of the scaled eigenvalues of $\mathcal{M}$ 
expressed as $r=(a^2+b^2)^{1/2}$ and $\theta=\cos^{-1}(\max(a,b)/r)$ with 
$a=\lambda^{\hat{\mathcal{M}}}_{\ne{}norm+1}$ and 
$b=\lambda^{\hat{\mathcal{M}}}_{\ne{}norm-1}$, \emph{i.e.} the other two 
eigenvalues not $\delta_{norm}$ (though these may be repeating)
and $y=\ln(\sin(2\theta))$, $x=\ln(r)$ with
\begin{align}
g_i  = & c_i(1) + c_i(2)x + c_i(3)y + c_i(4)x^2 + c_i(5)xy + c_i(6)y^2 + c_i(7)x^3  + c_i(8)x^2y + \nonumber \\
          & c_i(9)xy^2 + c_i(10)y^3 + c_i(11)x^4 + c_i(12)x^3y  +  c_i(13)x^2y^2 + c_i(14)xy^3 + c_i(15)y^4 \nonumber \\        
\label{gp}
\end{align}
were used.  Data used to fit $g$-coefficients was found from numerical integration of twice 
the gradient of the inertial range spectral energy density tensor integrated from $k_{min}$ to 
$k_c$ with $k_c$ corresponding to a wide range of filter anisotropies and a constant ratio of 
the coarsest filter width to $k_{min}$ of 32.  The coefficients in $g_i$ depend on the numerical 
method in use; filtering in the numerical data calculations was performed according.  We find
\vspace{-1cm}
\begin{multicols}{3}
\begin{align*}
   c_0(1)  &= -1.344 \nonumber \\
   c_0(2) &= -1.299 \nonumber \\
   c_0(3) &=  0.3625 \nonumber \\
   c_0(4) &= -0.01143 \nonumber \\
   c_0(5) &= -0.0002359 \nonumber \\
   c_0(6) &=  0.2315 \nonumber \\
   c_0(7) &=  0.001695 \nonumber \\
   c_0(8) &= -0.0002870 \nonumber \\
   c_0(9) &= -0.0009096 \nonumber \\
   c_0(10) &=  0.06177 \nonumber \\
   c_0(11) &= -0.00009189 \nonumber \\
   c_0(12) &=  0.00005443 \nonumber \\
   c_0(13) &=  0.0001554 \nonumber \\
   c_0(14) &=  0.0001159 \nonumber \\
   c_0(15) &=  0.005991 \nonumber \\
\end{align*}

\begin{align*}
   c_1(1) &=  1.953 \nonumber \\
   c_1(2) &=  0.03871 \nonumber \\
   c_1(3) &= -0.003524 \nonumber \\
   c_1(4) &= -0.009029 \nonumber \\
   c_1(5) &=  0.007419 \nonumber \\
   c_1(6) &=  0.001213 \nonumber \\
   c_1(7) &=  0.001038 \nonumber \\
   c_1(8) &= -0.001149 \nonumber \\
   c_1(9) &=  0.0008780 \nonumber \\
   c_1(10) &=  0.0006097 \nonumber \\
   c_1(11) &= -0.00004717 \nonumber \\
   c_1(12) &=  0.00006688 \nonumber \\
   c_1(13) &= -0.00004775 \nonumber \\
   c_1(14) &=  0.00005497 \nonumber \\
   c_1(15) &=  0.00007598 \nonumber \\
\end{align*}

\begin{align*}
   c_2(1) &= -0.01496 \nonumber \\
   c_2(2) &=  0.002111 \nonumber \\
   c_2(3) &= -0.005774 \nonumber \\
   c_2(4) &=  0.0002191 \nonumber \\
   c_2(5) &=  0.0006652 \nonumber \\
   c_2(6) &= -0.001771 \nonumber \\
   c_2(7) &= -0.00008604 \nonumber \\
   c_2(8) &=  0.000003282 \nonumber \\
   c_2(9) &=  0.0001955 \nonumber \\
   c_2(10) &= -0.0002762 \nonumber \\
   c_2(11) &=  0.000006200 \nonumber \\
   c_2(12) &= -0.000004179 \nonumber \\
   c_2(13) &= -0.000009257 \nonumber \\
   c_2(14) &=  0.00001253 \nonumber \\
   c_2(15) &= -0.00001806 \nonumber \\.
\end{align*}
\end{multicols}
The dynamic coefficient equation was altered to be an independent 
transport model for $\alpha$ as

\begin{equation}
\pd_t{}\alpha+\bar{u}_j\pd_j\alpha=\frac{1}{T_\alpha}\big{(}S_r-D_r-F_r\big{)}+\frac{S_c}{T_c}+\pd_k\big{(}\tfrac{\nu_t}{\sigma_\alpha}\pd_k{}\alpha\big{)}
\label{alpha}
\end{equation}
where
\begin{equation}
S_r = \left\{
\begin{array}{rl} \tanh(r_k-1)&\text{if }r_k\ge1,\\ 
                           \tanh\big{(}1-\tfrac{1}{r_k}\big{)} & \text{if } r_k < 1,
\end{array} \right.
\label{Sr}
\end{equation}
\begin{equation}
D_r=w_{rans}\max(\alpha+S_r-1,0),
\end{equation}
\begin{equation}
w_{rans}=\tanh\big{(}\tfrac{1}{2}\max(r_\varepsilon-1,0)^{1/4}\big{)},
\label{wrans}
\end{equation}
\begin{equation}
F_r=\min(\alpha+S_r,0),
\label{Fr}
\end{equation}
\begin{equation}
T_\alpha=C_{\alpha}\max\bigg{(}\frac{k}{\varepsilon},C_T\Big{(}\frac{\nu}{\varepsilon}\Big{)}^{1/2}\bigg{)},\\
\label{Ta}
\end{equation}
\begin{equation}
S_c = \left\{
\begin{array}{rl} 1&\text{if }S_r \text{ }\&\text{ }T_c\ge0,\\ 
                           0 & \text{otherwise},                           
\end{array} \right.
\label{Sc}
\end{equation}
\begin{equation}
T_c=\frac{L_m}{\max\big{(}\bar{u}_j\partial_j\mathcal{M}_{mn}\big{)}},
\label{Tc}
\end{equation}
and two parameters which compare the estimates of unresolved turbulent fluctuations and 
turbulent dissipation are respectively
\begin{equation}
r_k=C_{kQ}C_{sf}\frac{4}{9}\frac{k}{\overline{v^2}}\frac{\max(\zeta_{im}\mathcal{Q}_{mn}\zeta_{nj})}{\overline{v^2}},
\label{rk_v2f}
\end{equation}
and
\begin{equation}
r_\varepsilon=C_\varepsilon{}r_k^{3/2}\frac{L_{m}}{\delta_{max}},
\label{re}
\end{equation}
where $\delta_{max}$ is the magnitude of the projection of $\mathcal{M}$ on the direction 
of the maximum eigenvalue of $\zeta_{im}\mathcal{Q}_{mn}\zeta_{nj}$.  Coefficients were  
empirical tuned as $\sigma_\alpha=10$, $C_\alpha=4$, $C_{kQ}=8$, 
and $C_{\varepsilon}=0.03125$. Initial condition are set to basic RANS with $\alpha_o=1$ 
as well as wall and inlet boundary conditions $\alpha|_{wall}=1$, $\alpha|_{inlet}=1$, while 
outlet conditions are set to standard convective outflow, $\partial_n\alpha|_{outflow}=0$.   
Anisotropy in $\nu_{ij}$ was modified as  
\begin{equation}
a_{ij}=w_{RANS}\delta_{ij} + (1-w_{RANS})\sqrt{3}|\mathcal{Q}|^{-1}\mathcal{Q}_{ij}
\label{aij_m2}
\end{equation}
which was again scaled by $\nu_t$ and $\alpha$ to arrive at $\tau_{ij}$.
Model B still suffered from all the main shortcomings identified in Model A with the exception 
of correcting the tensor structure function for anisotropic resolution.  Though, it is worth 
pointing out that the ad-hoc RANS/LES blending in (\ref{wrans})  and (\ref{aij_m2}), when 
coupled with the anisotropy in $\mathcal{Q}_{ij}$ resulted in the nearly correct $C_f$ leading 
to the top of the hump (Fig. \ref{fig:hump_cpcf}).  This $C_f$ feature has not been captured 
in other simulations (\emph{e.g.} \cite{you:2006b}).  While it was no appropriate to the use grid-based 
length to formulate a RANS-region stress anisotropy, the grid changing lengthscale near the wall 
may have been mimicking the change in the turbulent lengthscale.  These results suggest 
anisotropy based on contraction of the resolved gradient-gradient product, with the correct 
turbulent lengthscale, may lead to improved basic RANS.\\*

\noindent\textbf{\emph{Model C:}}
The primary modification in Model C was a strict reinterpretation of the hybridization parameter as 
the ratio of modeled to total TKE, or $\alpha=k_{sgs}/k_{tot}$,  as in other models such as PANS
\cite{giri:2006}. 
Starting from a Deardorff $k_{sgs}$ transport equation,
\begin{equation}
D_tk_{sgs}=\alpha^2\mathcal{P}_{sgs}-\varepsilon+\partial_i(\alpha^2\nu_t\partial_ik_{sgs}),
\end{equation}
where $\nu_t=C_{\mu}k_{tot}^2/\varepsilon$, $\mathcal{P}_{sgs}=2\alpha^2\nu_t\overline{S}_{ij}\overline{S}_{ij}$, 
the supposedly equivalent $k_{tot}$ equation was derived with $D_tk_{sgs}=D_t(\alpha{}k_{tot})$ as 
  \begin{equation}
  D_tk_{tot}=\alpha\mathcal{P}_{sgs}-\frac{k_{tot}}{\alpha}D_t\alpha-\frac{\varepsilon}{\alpha}+\frac{1}{\alpha}\partial_i\big{(}\nu_t\partial_i(\alpha{}k_{tot})\big{)}.
  \end{equation}
The evolution of the level of resolved turbulence was modeled as in Model B (\ref{alpha}).  Further, 
the resolution adequacy parameter was changed to the legthscale-based approach as in (\ref{rd}).  
Again, this method failed to recognize the deleterious effects of including fluctuating quantities in non-linear 
RANS source terms as well as using the standard merging mean and fluctuating contributions to $\tau_{ij}$ 
with a single eddy viscosity.  Further, as this model explicitly included a sink term in $k_{tot}$ due to 
the total change in $\alpha$, it implicitly assumed that precise amount of resolved $k$ was being 
added.  This is, of course, not possible when relying on passive self-generation.\\*

\noindent\textbf{{II. Current model details}}\\
For completeness, the models used in the presented results are provided here.  
The SGS portion of the MS formulation makes use of the ``code-friendly'' 
version~\cite{lien:2001} of Durbin's $\overbar{v^2}$-$f$ model~\cite{durb:1995}.  
The three associated equations for turbulent kinetic energy, $k$, turbulent 
dissipation rate, $\varepsilon$, and minimum turbulent stress component, 
$\overbar{v^2}$, take the generic form
\begin{equation}
D_t\phi = \mathcal{P}_\phi-\mathcal{D}_\phi+\mathcal{V}_\phi
\end{equation}
where
\begin{equation}
\mathcal{V}_\phi=\partial_k\big{(}(\nu+\sigma_{\phi}\nu_t) \partial_k\phi\big{)}.
\end{equation}
With the eddy viscosity as $\nu_t=C_\mu\overbar{v^2}T$, source terms, wall boundary 
conditions, and diffusion coefficients are as follows\\
\vspace{-1cm}
\begin{multicols}{3}
\begin{align*}
\mathcal{P}_k &= 2\nu_tS^2\\
\mathcal{D}_k &= \varepsilon\\
\sigma_k &= 1\\
k_{wall} &= 0
\end{align*}

\begin{align*}
\mathcal{P}_{\varepsilon} &= \frac{\mathcal{P}_k}{T}\\
\mathcal{D}_{\varepsilon} &= \frac{\varepsilon}{T}\\
\sigma_{\varepsilon} &= {1/1.3}\\
\varepsilon_{wall} &= 2\nu\frac{k^2}{\delta^2_{wall}}
\end{align*}

\begin{align*}
\mathcal{P}_{v2} &= kf\\
\mathcal{D}_{v2} &= 6\frac{\overbar{v^2}}{k}\varepsilon\\
\sigma_{v2} &= 1\\
\overbar{v^2}_{wall} &= 0
\end{align*}
\end{multicols}
where $S=(\la{}S_{ij}\ra\la{}S_{ij}\ra)^{1/2}$, $\delta_{wall}$ is the distance to the first cell center, and the turbulence time scale is
\begin{equation}
T = \min\bigg{(} \max\Big{(}\frac{k}{\varepsilon},6\frac{\sqrt{\nu}}{\varepsilon} \Big{)}, \frac{0.6k}{\sqrt{6}C_\mu\overbar{v^2}S}\bigg{)}.
\end{equation}
The model for the redistribution rate, $f$, appearing in the production of $\overbar{v^2}$ is of the Helmholtz type
\begin{equation}
(C_LL)^2\partial_k\partial_k{}f-f=R_f\bigg{(}\frac{\overbar{v^2}}{k}(C_1-6)-\frac{2}{3}(C_1-1)\bigg{)}-C_2\frac{\mathcal{P}_k}{k}
\end{equation}
where
\begin{equation}
L= \max\bigg{(} \min\Big{(}\frac{k^{3/2}}{\varepsilon},   \frac{k^{3/2}}{\sqrt{6}C_\mu\overbar{v^2}S} \Big{)},  C_\eta\frac{\nu^{3/4}}{\varepsilon^{1/4}} \bigg{)}
\end{equation}
and an additional time scale modification is made here with 
\begin{equation}
R_f=\min\Big{(} \frac{1}{T}, \frac{S}{3}\Big{)}
\label{Rf}
\end{equation}
which was found to improve behavior for higher $Re$ channel flow.  Without this modification, $\overbar{v^2}$ 
becomes excessive in hybrid simulations in response to $k$ increasing towards the DNS value 
while the basic RANS $\overbar{v^2}$ is already near the DNS value.  Due to how $r_\mathcal{M}$ is formulated 
based on $\overbar{v^2}$, this error resulted in forcing being activated too near the wall.  Therefore, when not using 
the modification in (\ref{Rf}), additional near-wall protection is necessary.   An additional scaling-factor of 
\begin{equation}
\eta_{wall}=\frac{1}{2}\Big{(}\tanh\big{(} C_\zeta(\zeta-\zeta_c)\big{)} +1 \Big{)}
\end{equation}
applied to (\ref{Fi_actual}), with $\zeta=1.5\overbar{v^2}/k$, was found to be sufficient.  The model is complete with 
the following coefficients
\begin{align*}
C_\mu &= 0.2, { } C_1 = 1.4, { } C_2 = 0.3, C_L = 0.23,\\
C_{\varepsilon{}1} &= 1.4\Big{(}1+0.005(k/\overbar{v^2})^{1/2} \Big{)}, { } C_{\varepsilon{}2} = 1.9, \\
C_\eta &= 70, {} C_\zeta=10, {}\zeta_c=0.55
\end{align*}


\noindent\textbf{\emph{M43 SGET:}}
The M43 model form is provided in (\ref{nu_m43}).  Here, $C(\mathcal{M})$ is 
provided.  Again, using the polynomial form of (\ref{gp}), coefficients for $2^{nd}$-order 
finite volume numerics are 
\begin{align*}
c=(1.034,
-0.1541,
-0.007737,
0.1776,
0.06087,
0.1622,
-0.04109,
-0.02738,\nonumber \\
0.005521,
0.04914,
0.002926,
0.002673,
0.0004864,
0.002136,
0.005113)
\end{align*}
with $C^\circ_\mathcal{M}=0.13$.   Naturally, the use of polynomials to fit the 
coefficients limits their applicability to the range of fitting data.  The above 
coefficients make us of data up to ($128:1$) aspect ratio cells.  Further, $C(\mathcal{M})$ 
is scaled by $\max{}\big{(}\min{}(\la{}r_\mathcal{M}\ra,10),1\big{)}$ to rapidly 
remove resolved fluctuations  where resolved turbulence has been transported into 
regions with insufficient resolution.  This modification is essentially ad-hoc and 
methods of removing it will be studied in future work.\\

\vspace{1cm}

\section*{Acknowledgements}
This material is based upon work supported by NASA under cooperative
agreement number NNX15AU40A.

\bibliography{paper}

\end{document}